\documentclass{emulateapj}
\usepackage{amsmath}

\newcommand{\msun}{\,M_\odot}
\newcommand{\hmsun}{\,{\it h}^{-1}M_\odot}
\newcommand{\kms}{\,\rm{km~s}^{-1}}
\newcommand{\mrei}{M_{\rm{rei}}}

\newcommand{\msat}{M_{\rm{sat}}}
\newcommand{\Msat}{M_{\rm{sat}}}

\newcommand{\mtot}{M_{\rm{tot}}}
\newcommand{\mthree}{M_{\rm 300 }}
\newcommand{\zrei}{z_{\rm{rei}}}
\newcommand{\zsat}{z_{\rm{sat}}}
\newcommand{\vcrit}{V_{\rm{crit}}}
\newcommand{\vcirc}{V_{\rm{circ}}}
\newcommand{\vcritr}{V_{\rm{crit,r}}}
\newcommand{\tvir}{T_{\rm vir}}
\newcommand{\K}{\,{\rm K}}
\newcommand{\etal}{et al.\ }
\newcommand{\Nobs}{N_{\rm obs}(M_V)}

\begin{document}
\title{A quantitative explanation of the observed population of Milky Way
satellite galaxies.}
\shorttitle{Milky Way satellite galaxies.}

\author{
	Sergey E. Koposov\altaffilmark{1,2,3},
	Jaiyul Yoo\altaffilmark{4},
	Hans-Walter Rix\altaffilmark{1},
	David H. Weinberg\altaffilmark{5},
	Andrea V. Macci\`o\altaffilmark{1},
	Jordi Miralda-Escud\'e\altaffilmark{6},
       }

\altaffiltext{1}{Max Planck Institute for Astronomy, K\"{o}nigstuhl 17, 69117
Heidelberg, Germany; \email{koposov@mpia.de}}
\altaffiltext{2}{Institute of Astronomy, University of Cambridge, Madingley
Road, Cambridge, UK}
\altaffiltext{3}{Sternberg Astronomical Institute, Universitetskiy pr. 13,
119992 Moscow, Russia}
\altaffiltext{4}{Harvard-Smithsonian Center for Astrophysics, Harvard
University, 60 Garden Street, Cambridge, MA 02138}
\altaffiltext{5}{Ohio State University, Department of Astronomy and CCAPP, 
Columbus, OH 43210, USA}
\altaffiltext{6}{Instituci\'{o} Catalana per la Recerca i Estudis Avan\c{c}ats,
Barcelona, Spain / Institut de Ci\`{e}ncies del Cosmos, Universitat
de Barcelona, Spain}

\begin{abstract}
We revisit the well known discrepancy between the observed number of Milky 
Way (MW) dwarf satellite companions and the predicted population of cold dark 
matter (CDM) sub-halos, in light of the dozen new low luminosity satellites 
found in imaging data from the Sloan Digital Sky Survey (SDSS) and our recent 
calibration of the SDSS satellite detection efficiency, which implies a total 
satellite population far larger than these dozen discoveries. We combine a 
detailed dynamical model for the CDM sub-halo population with simple, 
physically
motivated prescriptions for assigning a stellar content to each sub-halo, then
apply observational selection effects and compare to the current observational
census. Reconciling the observed satellite population with CDM predictions 
still requires strong mass-dependent suppression of star formation in low mass 
sub-halos: models in which the stellar mass is a constant fraction 
$F_*(\Omega_b/\Omega_m)$ of the sub-halo mass $M_{\rm sat}$ at the time it 
becomes a satellite fail for any choice of $F_*$. However, previously advocated
models that invoke suppression of gas accretion after reionization in halos 
with circular velocity $\vcirc \leq \vcrit \approx 35\kms$ can reproduce the 
observed satellite counts for $-15 \leq M_V \leq 0$. Successful models require 
$F_* \approx 10^{-3}$ in halos with $\vcirc > \vcrit$ and strong suppression 
of star 
formation {\it before} reionization in halos with $\vcirc \la 10\kms$; models 
without pre-reionization suppression predict far too many satellites with 
$-5\leq M_V \leq 0$. In this successful model, the dominant fraction of
stars formed after reionization at all luminosities. Models that match the
satellite luminosity distribution 
also match the observed heliocentric radius distribution, and they reproduce 
the observed characteristic stellar velocity dispersion 
$\sigma_* \approx 5 -10\kms$ of the SDSS dwarfs given the observed sizes 
($\sim 50-200\,$pc) of their stellar distributions. The model satellites have 
$M(<300\,{\rm pc})\sim 10^7 M_\odot$ as observed even though their present 
day total halo masses span more than two orders of magnitude; the constancy of 
central masses mainly reflects the profiles of CDM halos. Our modeling shows 
that natural physical mechanisms acting within the CDM framework can 
quantitatively explain the properties of the MW satellite population as it is 
presently known, thus providing a convincing solution to the `missing 
satellite' problem.
\end{abstract}

%\maketitle
\section{Introduction}\label{sec:intro}

The inflationary cold dark matter scenario predicts an initial
fluctuation spectrum with power that continues down to 
small scales, and in consequence it predicts a mass function
of dark matter halos that rises steeply towards low masses.
A significant fraction of these halos survive as gravitationally self-bound
units long after falling into more massive halos.
As pointed out forcefully by \citet{klypin99} and
\citet{moore99}, the predicted number of sub-halos within a
Milky Way-like galaxy halo greatly exceeded the then known
numbers of Milky Way or Local Group dwarf satellites, when sub-halos
and observed dwarfs were matched based on velocity dispersion
or corresponding circular velocity \citep[see also][]{kauffmann93}.  This
discrepancy between
predicted and observed numbers has become known as the
``missing satellite problem.''

Proposed solutions fall into three general categories.
The first modifies the properties of dark matter or the
primordial fluctuations from inflation in a way that 
eliminates the low mass dark matter sub-halos themselves
\citep[e.g.][]{kamionkowski00,spergel00,bode01,zentner03}.
The second appeals to astrophysical mechanisms that suppress
star formation in low mass halos so that they do not
become observable dwarf satellites; photo-heating by
the meta-galactic UV background is an attractive mechanism
because it naturally introduces a cutoff at approximately
the correct velocity scale \citep{bullock00,somerville02,kravtsov04}.
The third possibility, arguably a variant of the second, is that
the numerous dwarf companions of the Milky Way actually exist but have been
missed by observational searches.

In this paper we revisit the ``missing satellite problem''
with particular emphasis on the role of the new dwarf
companions discovered in imaging data from the 
Sloan Digital Sky Survey (SDSS; \citealt{york00}; \citealt{adelman08}).
There are now about a dozen of these
(\citealt{willman05,belokurov06,belokurov07,zucker06,irwin07,koposov07,
walsh07}; a couple of systems still have ambiguous status), most of them
at least an order of magnitude less luminous than the faintest of the
previously known, ``classical'' satellites.\footnote{Throughout the paper we use
``faint'' and ``bright'' to refer to intrinsic luminosity rather than apparent
brightness.}
Spectroscopic follow-up
\citep[e.g.][]{martin07,simon07,geha08} for many of them indicates that 
they are indeed dark matter
dominated systems, even though most are fainter than typical globular
clusters, as low as only $\sim $1000 L$_\odot$
\citep[e.g.][]{belokurov07,martin08}. Remarkably, almost all 
of the newly found faint
satellite galaxies have stellar velocity dispersions in the range $3-10\kms$,
though their luminosities vary widely. 
Similarly, the total masses within the inner 300 pc span less
than an order of magnitude \citep{strigari08}.

Since the SDSS imaging in which these satellites have been discovered
covers only $\sim 20\%$ of the sky, a naive accounting would
increase the estimated number of Milky Way companions by $5\times 12 = 60$,
in addition to the ten classical satellites.
However, \cite{lfpaper} use a well-defined identification algorithm
to show that the SDSS dwarfs are also subject to strong radial
selection effects.  Most of the newly discovered objects could
only have been found within distances of 50-100 kpc, much smaller
than the inferred virial radius of the 
Milky Way's dark matter halo
($\sim 280$ kpc for $\rho_{\rm vir}/\bar{\rho} = 340$; \citealt{xue08}).
The faintest SDSS dwarfs are detectable over only 1/1000 of the
halo virial volume (including the factor of five for sky coverage).
\citet{walsh08} have recently reached similar conclusions 
based on an independent identification algorithm and independent
Monte Carlo tests.

Such analyses are the basis for `volume corrections' for the faint Milky Way
satellite population. With proper volume corrections applied, the luminosity
function of faint Milky
Way satellite galaxies turns out to be a rather shallow power law in the range
$-15<$M$_V < -3$ \citep{lfpaper}.
These results in turn imply that
the number of satellites brighter than $M_V=-3$ is
$\sim$ 80 or more, and the number above $M_V=0$ could be
a few hundred.  \cite{tollerud08} reached a similar conclusion,
adopting a radial satellite distribution
based on the {\it Via Lactea} simulation of \cite{diemand07}.
Even this census counts only dwarfs that are above the effective
surface brightness threshold for SDSS detection.
With the \cite{lfpaper} detection algorithm, this threshold is
approximately 30 mag arcsec$^{-2}$ ($V$-band), with a weak 
dependence on luminosity and distance.
The dwarfs found in SDSS have surface brightnesses that
range from 24 to 30 mag arcsec$^{-2}$.

Studies of the high redshift Ly$\alpha$ forest indicate that the
small scale power expected in the standard $\Lambda$CDM scenario
(inflationary cold dark matter with a cosmological constant)
is indeed present in the primordial fluctuation spectrum
\citep{narayanan00,viel05,abazajian06,seljak06}.  Astrophysical suppression of
star 
formation, and photo-ionization suppression in particular,
has emerged as the most plausible and hence popular solution to the ``missing
satellite''
conundrum.  Within this category, there have been different
proposals about what sub-halos host the observed dwarf satellites.
\cite{bullock00} suggested that the observed dwarfs are those
whose sub-halos assembled a substantial fraction of their
mass before reionization, and thus before the onset of
photo-ionization suppression.  \cite{stoehr02} suggested that
the measured stellar velocity dispersions are well below the
virial velocity dispersions of the dark matter sub-halos,
and that the observed dwarfs occupy sub-halos that are still
above the velocity threshold where star-formation suppression occurs.
\cite{kravtsov04} used N-body simulations to show that roughly 10\%
of sub-halos lose a large fraction ($\sim 90\%$) of their mass
during dynamical evolution without being completely disrupted;
they suggested that the observed dwarfs occupy sub-halos that
were above the suppression threshold at the time they became
satellites but have suffered extensive mass loss since then.
These papers and others (e.g., \citealt{somerville02,strigari07,orban08})
focus on explaining the classical (pre-SDSS) dwarf spheroidal
population, with luminosities in the range $-8 < M_V < -15$ (excluding
the Magellanic clouds) and stellar velocity dispersions
in the range $8\kms < \sigma_* < 25\kms$.
The  recently discovered SDSS dwarfs have much lower luminosities 
($-8<M_V<-1.5$), lower surface brightness,
and somewhat lower velocity dispersion
($\sigma_* \sim 5 \kms$), so they could have a distinct
formation mechanism, or they could form a continuum
with the classical dwarf spheroidals.
%[Insert a section about the Vcirc threshold before the reionization]

The new SDSS discoveries and their quantified detectability are the basis for
the model-data comparison in this paper.
We construct and test models of
the Milky Way dwarf satellite population that incorporate Monte Carlo
realizations
of merger trees for $10^{12} M_\odot$ (main galaxy) halos, 
a detailed analytic model for the dynamical evolution
and disruption of sub-halos, and a variety of recipes for
assigning stellar masses to these sub-halos motivated by
ideas in the existing literature.  For most of our models,
we assume that a sub-halo can only accrete gas to form
stars (a) before the epoch of reionization or (b) after 
reionization if its virial velocity exceeds a critical 
threshold before it enters the Milky Way halo and becomes
a satellite.  The spirit of the exercise is similar to that
of \cite{bullock00}, but the dynamical modeling of 
sub-halos is more sophisticated, and we are now in a position to include
directly
the (strong)
constraints imposed by the SDSS dwarfs accounting for the radial
selection function found by \cite{lfpaper}.  
In contrast to most previous studies, we treat the luminosity
distribution as the primary test of models, rather than
the stellar velocity dispersions or central masses
\citep{strigari07,strigari08,li08,maccio08},
or the inferred but unobservable sub-halo circular velocities. 
This emphasis is
motivated by the fact that the luminosity is the
foremost quantity that matters for the observational selection. We consider
stellar velocity dispersions and central masses as an
additional test, but their interpretations are affected by the uncertainty in 
the dark matter profiles of the sub-halos associated with 
observed dwarfs.

\section{The Population of Dark Matter Sub-Halos in the Milky
Way}\label{sec:dm_halos}

  Our model for the Milky Way satellites is based on the cold dark matter
scenario, with each satellite forming initially in a separate dark matter halo
that at some point falls into the Milky Way's dark matter halo. 
We refer to the bound dark matter satellites orbiting 
in the Milky Way halo as sub-halos. A
sub-halo may or may not correspond to a dwarf satellite galaxy,
depending on whether it contains an observable number of stars. 
In this Section we describe
our model for computing the dynamical evolution of sub-halos.

We use the dynamical dark-matter-only model of sub-halos developed by
\citet{yoo07} 
to compute the sub-halo population and its orbital distribution.
This model is described in
detail in \citet{yoo07}, where a much larger halo of $10^{15}\msun$ was
considered as a model of a massive cluster of galaxies. Here we consider
instead a final halo of $10^{12}\msun$ at the present time as a
representation of the Milky Way galaxy. Despite the change in the final
halo mass, the model remains basically the same as described in
\citet{yoo07}, so here we make only a brief summary of its description.

  The model uses the extended Press-Schechter formalism to generate a
Monte Carlo merger tree of the parent halo at the present time
\citep{ps,bcek}. We follow the dynamical evolution of all the sub-halos
with masses $M_h > 10^6 \msun$ until they merge with the
Milky Way and lose their mass below $M_h=10^5 \msun$. All halos start as
isolated objects, and they grow in mass
by accretion and mergers for as long as they remain isolated. At some
redshift, $z_{\rm sat}$, they merge into a larger halo (either the Milky Way
or another object that will become a Milky Way sub-halo). After this
merger, the object has become a satellite or sub-halo and it stops
growing in mass. It can subsequently lose mass by tidal
stripping when it passes near the center of its parent halo or
undergoes encounters with other sub-halos. The sub-halo is subject to
dynamical friction, which tends to shrink its orbit, and to random
encounters with other sub-halos, which on average expand the orbit.
The orbital eccentricity is also subject to random variations. The model
allows for the presence of sub-halos within other sub-halos. When a
sub-halo is disrupted, any sub-halos it contained are dispersed into the
new, larger parent halo. This simple analytic model is able to reproduce
the sub-halo mass function, in reasonably good agreement with that found
in numerical $N$-body simulations \citep{zentner,shaw,yoo07}. For the
present purpose, this approach has the advantage (over N-body) of easily
affording the required mass resolution and multiple halo realizations.

  We adopt a flat $\Lambda$CDM cosmology with matter density
$\Omega_m=0.24$, baryon density $\Omega_b=0.04$, 
power spectrum normalization $\sigma_8=0.8$, Hubble constant $h=0.7$, 
and a primordial spectral index $n_s=0.95$, consistent with recent 
measurements (\citealt{wmap,max2}). The matter power spectrum is
computed by using the transfer function of \citet{eis}.
We generate six Monte Carlo merger trees of a Milky-Way sized halo. Each
realization provides the sub-halo mass function, their orbital elements
and density profiles at the present time. Our statistical results are the 
average of the six different realizations. 

  The dynamical model of Yoo et al.\ (2007) uses the Jaffe profile and
its velocity dispersion to model sub-halos and their dynamical
interactions, for reasons of numerical simplicity and because large
galaxies that are tidally-limited satellites of a larger halo are
reasonably well modelled by a Jaffe sphere for their baryon plus dark
matter density profiles. However, the very low-mass dwarf satellites
tend to be dominated by dark matter even in their inner parts. We
therefore make an adjustment to better connect our Monte Carlo
simulation results to the observed Milky Way dwarf galaxies:
we use the sub-halo masses and orbital elements,
which are the quantities most robustly computed in the Yoo
\etal (2007) model, but we calculate the density profiles and
velocity dispersions of sub-halos assuming that they have an
NFW profile \citep{navarro97}.  Using the standard spherical collapse
model, the virial radius of an isolated halo is assigned as
\begin{equation}
R_{\rm vir}= \left[ \frac{3\, M_{\rm halo}}{4\, \pi \,\Delta_c \,
\bar{\rho}_m(z)} \right]^\frac{1}{3} ~,
\end{equation}
where $\Delta_c=(18\,\pi^2+82\,x-39\,x^2)/(1+x)$ \citep{bryan98},
$x=-(1-\Omega_m)/(\Omega_m(1+z)^3+1-\Omega_m)$, and the mean cosmic
density is $\bar{\rho}_m(z)=\Omega_m\rho_c(1+z)^3$.
The halo concentration $c$ is computed using the relation from
\citet{bullock01}, scaled to $\sigma_8=0.8$ according to~\cite{maccio07},
with 
$c=0.8\times 9 \times \left(M_h/10^{13}h^{-1}M_\sun\right)^{-0.13}/(1+z)$.

For the model in this paper, we use in particular the sub-halo
masses at two different special epochs: $\mrei\equiv\mtot(z=\zrei)$
when the universe reionizes and the 
photo-ionization background starts to suppress the star formation efficiency
in low mass halos, and $\msat\equiv\mtot(\zsat)$ at the epoch when
a halo merges into a larger halo and we presume that subsequent
star formation and gas accretion is halted in the sub-halo. We shall
also use below the halo circular velocity $\vcirc$, which is the
virial circular velocity, $\vcirc\equiv [(G\,\mtot)/R_{\rm vir}]^{1/2}$.
(Here $\mtot$ refers to the total mass including dark matter and
a universal fraction of baryons.)

\begin{figure*}
\includegraphics[width=0.33\textwidth]{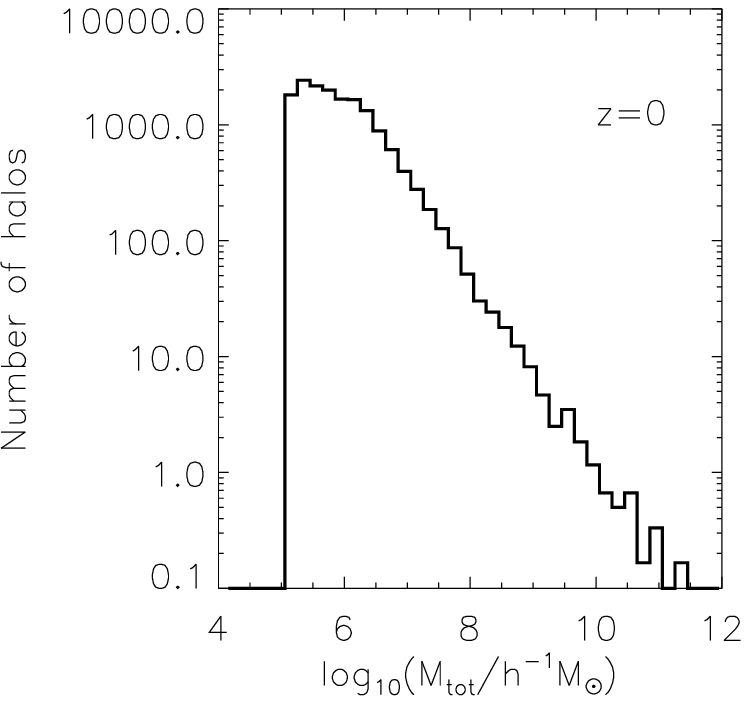}
\includegraphics[width=0.33\textwidth]{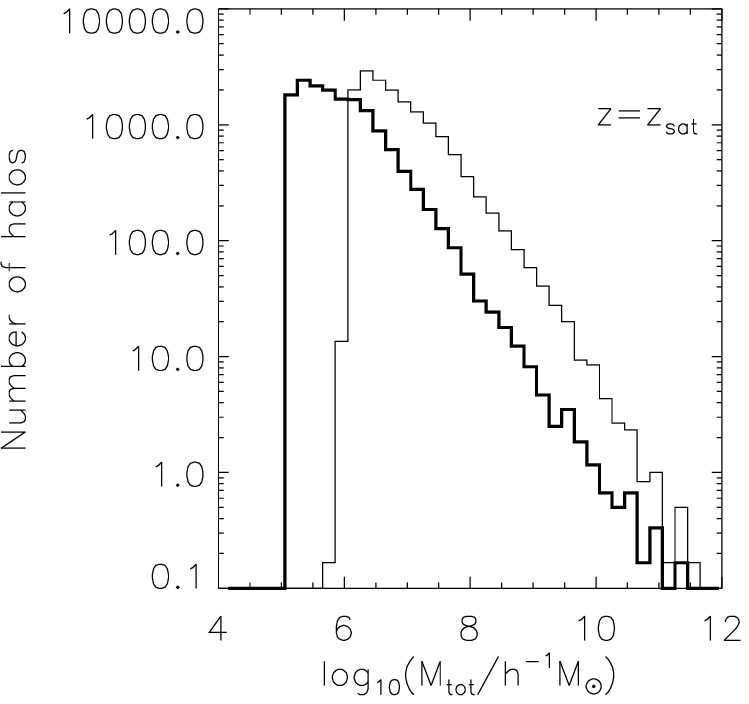}
\includegraphics[width=0.33\textwidth]{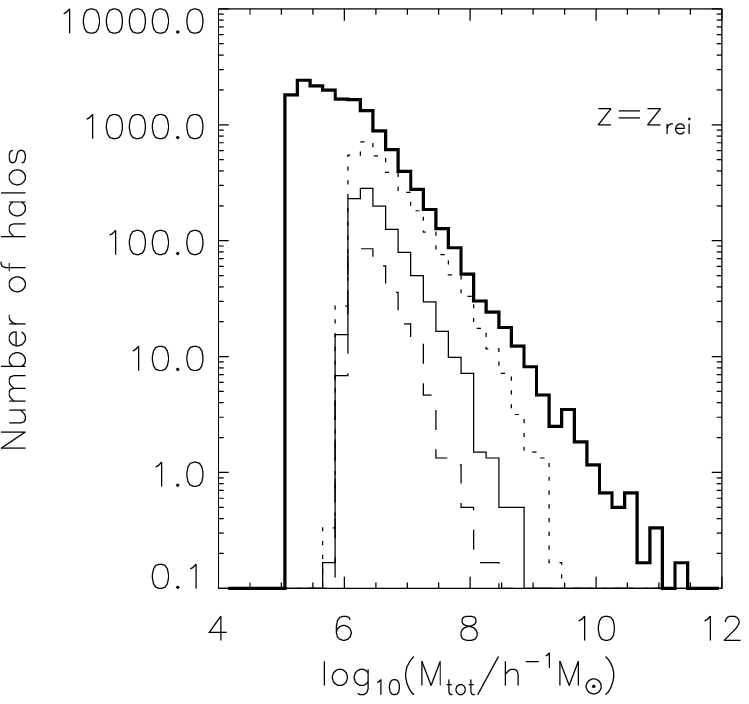}
\caption{
Mass distribution of dark matter sub-halos at different epochs.
{\it Left panel}: The present-day ($z=0$) mass function of sub-halos
within a Milky Way-like halo.  This histogram is repeated in the
other two panels for reference.
{\it Middle panel}: 
Distribution of mass that present-day sub-halos 
had at $z=\zsat$, the epoch at which they became
a satellite within a larger halo ({\it thin line}); tidal stripping of
satellite halos is manifesting important. 
{\it Right panel}: Mass distribution of present day MW sub-halos
at the epoch of reionization, for $\zrei=8$ ({\it dotted}), 
11~({\it thin solid}), and 14~({\it dashed}).
All panels reflect the
average of six different realizations of MW-like halos. 
The flattening below $M=10^6\hmsun$ and the
sharp cut-off at $M=10^5\hmsun$ arise from the mass 
resolution limits of our simulations.
}
\label{fig:halo_mass_distr}
\end{figure*}

\begin{figure}
 \includegraphics[width=0.5\textwidth]{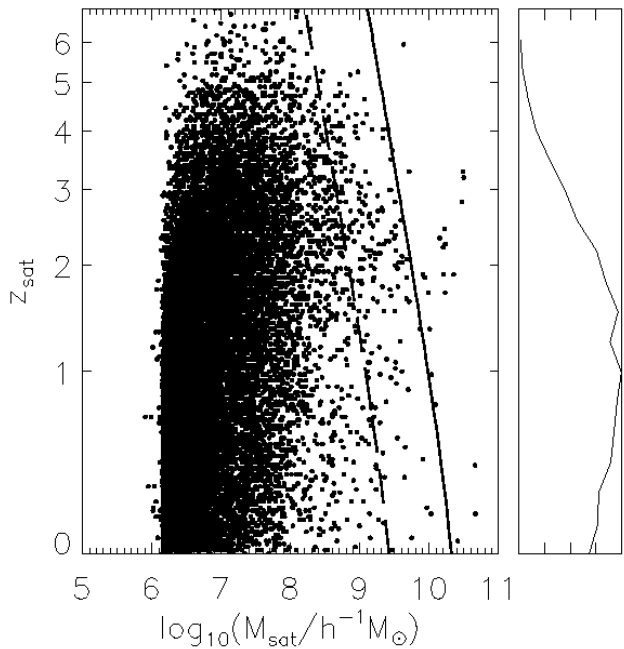}
\caption{Epochs when (sub-)halos were accreted into larger halos, and
masses at that time. This Figure illustrates the results from 
one Monte-Carlo realization
of the semi-analytic model, with each point showing
the redshift $\zsat$ at which a sub-halo first became 
a satellite in a larger halo
against its total mass $\msat$ at that epoch.
The small panel on the right shows the distribution of $\zsat$.
Solid and dashed lines show the locus of halos with 
$\vcirc(\zsat)=40\kms$ and $20\kms$, respectively.
}
\label{fig:accretion_history}
\end{figure}

In Figure~\ref{fig:halo_mass_distr} we show the distribution of
M$_{\rm{sub-halo}}$ at redshifts $z=0$ (left panel), $z=\zsat$ (middle
panel) and $z=8$, 11, and 14 (right panel). As expected, the mass distribution
is close to a power-law, except near the resolution limit of our
simulations. 
Also, in Figure~\ref{fig:accretion_history} we show the accretion history of
the MW sub-halo population, by plotting the halo masses of the present day MW
sub-halos at the time of accretion vs. the redshift at which they became
satellites of larger halos. We see that most of the MW sub-halos became
satellites at $z < 2$. 
Most of the accreted satellites have small circular velocities
$\vcirc<20 \kms$, so they lie in a range where gas accretion and
star formation are likely to be suppressed after the epoch
of reionization \citep{quinn96,thoul96,bullock00}.

\section{Populating the DM halos with Stars}
\label{sec:populating_halos}
\subsection{Recipes to assign stellar masses to sub-halos}
\label{sec:recipes} 
To make direct observational predictions from
these models, we populate each sub-halo in a given Monte-Carlo realization
with stars according to a sequence of recipes, then test how many
of these satellites could have been found within the SDSS. 
Some of these recipes are mathematically simple illustrations,
while others are motivated by the expected effects of ionization and 
cooling physics as discussed in the introduction.
For reference, the nomenclature of the
recipes is summarized in Table~\ref{tab:models}.
In all cases we calculate the stellar mass based on the sub-halo
mass (dark matter plus baryons in the universal fraction)
at the accretion epoch $\zsat$, denoted $\msat$.
We implicitly assume that satellites do not accrete new material
to form additional stars and that tidal stripping of the dark
matter does not affect the stellar content of the satellite
if it survives to the present day.
Simulations suggest that these assumptions are reasonable but
not perfect approximations \citep{simha08,penarrubia08}.

We begin with the simplest model (denoted Model 1A), that the 
stellar mass is a constant fraction of the sub-halo mass at the time of
accretion into the main halo:
\begin{equation}
M_*=f_*\times \msat.
\end{equation}
The arguments of \cite{klypin99} and \cite{moore99} suggest that this
model will fail badly, and we show that it does indeed fail despite
the new satellite discoveries and the radial selection biases that
affect them.  There is ample evidence that the efficiency
of star formation declines rapidly towards low masses even well
above the dwarf satellite regime (e.g., \citealt{vandenbosch07}).
In Model 1B, we allow the stellar fraction to vary as a power law
of $\msat$ below a threshold $M_0$:
\begin{equation}
M_* = f_* \times {\tt min}\left(\left(\frac{\msat}{M_0}\right)^\alpha,1\right)
  \times \msat~.
\end{equation}

Our second approach to modeling stellar masses 
includes the effects of a pervasive energetic radiation
field after the epoch of reionization, which heats gas and hence
keeps it from accumulating at the centers of low-mass halos.
Calculations by \cite{quinn96} and \cite{thoul96} 
showed that gas accretion in  halos with
the circular velocities below $\vcirc\sim 30-40\kms$
is strongly suppressed, while substantially larger halos are
minimally affected
(see also \citealt{weinberg97,gnedin00}).
In this spirit, we assume that halos below a critical circular
velocity form no stars after reionization, and we thus assign stellar masses
\begin{equation}
M_*=
\begin{cases}
    f_*\times \msat & \text{if }\vcirc(\zsat)>\vcrit \\
    f_*\times \mrei & \text{if
}\vcirc(\zsat)<\vcrit.
\end{cases}
\label{eq:sharp_vcrit_model}
\end{equation}
This model (Model 2) has three adjustable parameters ---
$f_*$, $\vcrit$, and $\zrei$ --- with expectations that 
$\vcrit \sim 20-40\kms$ and 
$\zrei\sim 11$ \citep[e.g.][]{weinmann07,dunkley08}. 
The approach is similar to that of 
 \citet{bullock00}, except that we treat $\vcrit$ as free
parameter, and the stellar mass formed before the epoch of reionization is
assigned using $M_*=f_*\times \mrei$, instead of simply dividing
galaxies into
``observable'' or ``unobservable'' classes based on the fraction of the mass
accreted by $\zrei$. 

Our third class of models is similar to the second, but it replaces
the sharp threshold of equation~(\ref{eq:sharp_vcrit_model}) with
the continuous transition found in numerical simulations by
\cite{gnedin00}, \cite{hoeft06}, and \cite{okamoto08}.
The numerical results in these papers can be described fairly
well by a formula similar to that in \cite{gnedin00}, with the
fraction of baryons that cool in low mass halos suppressed by a factor
$[1+0.26(\vcrit/\vcirc)^3]^{-3}$; well after the reionization
redshift, the critical velocity is found to be approximately
independent of redshift. \cite{gnedin00} found $\vcrit \sim 40\kms$,
but these results were artificially affected by numerical resolution
(N. Gnedin, private communication).  \cite{hoeft06} and \cite{okamoto08}
find $\vcrit \sim 25-30\kms$.
Including the pre-reionization contribution to $M_*$, this model
(Model 3A) becomes
\begin{equation}
M_* = \frac{f_* \times
(\msat-\mrei)}{(1+0.26\,(\vcrit/\vcirc(\zsat))^3)^3}+f_*\times\mrei ~.
\label{eq:gnedin_recipe_vc}
\end{equation}

The assumption that all halos can form stars
before $\zrei$ may not be justified because in halos with virial
temperature $\tvir \la 10^4\K$ ($\vcirc \la 10\kms$) the gas does
not get hot enough to cool by atomic processes, and simulations
that include molecular cooling suggest that gas cooling and
star formation is very inefficient in such halos
\citep{haiman97,barkana99,machacek01,wise07,oshea08,bovill08}. 
We will therefore consider variant models (Model 3B) that eliminate
stellar mass in pre-reionization halos below a critical
threshold $\vcritr\sim 10 \kms$.\footnote{We will refer to these
as models with ``pre-reionization suppression,'' but this simply 
means that halos with $\vcirc(\zrei)$ below a critical threshold
form stars with very low efficiency (too low to produce observable
satellites), most likely because of inefficient cooling rather than
active feedback.}
In Model 3B, halos with $\vcirc(\zrei) < \vcritr$ 
have stellar mass
\begin{equation}
M_* = \frac{f_* \times \msat}{(1+0.26\,(\vcrit/\vcirc(\zsat))^3)^3}~,
\end{equation}
while halos with $\vcirc(\zrei) > \vcritr$ have mass given by 
equation~(\ref{eq:gnedin_recipe_vc}).

To determine very roughly the plausible range of values for
the stellar mass fraction $f_*$, we can refer to the results of 
\citet{strigari07}, who derived $M(<r_{\rm tidal})/L)=30-800 M_\odot/L_\odot$
for the classical dwarfs, and \citet{simon07}, who measured
velocity dispersions for SDSS dwarfs and inferred total mass-to-light
ratios of $140-1800 M_\odot/L_\odot$. 
For a stellar mass-to-light ratio $M_*/L_V=1 M_\odot/L_\odot$,
we infer plausible values of $f_* \sim 10^{-4} - 10^{-2}$,
though these are very uncertain because all the dynamical
mass-to-light ratio determinations suffer from the fact that the stars in
luminous bodies of the dSphs probe only the inner parts of 
the dark matter potential wells. 
Another line of argument comes from matching the mean space density
of dark matter halos to that of observed {\it field} dwarfs:
\cite{tinker08} find $f_* \approx 10^{-3.6}$ at absolute
magnitude $M_r \approx -10$.
In the rest of the paper, we will frequently refer to the stellar
mass fraction normalized by the universal baryon fraction:
\begin{equation}
F_*\equiv \frac{f_*}{\Omega_b/\Omega_m} = 6.25 f_*.
\end{equation}
Note that $f_*$ and $F_*$ refer to stellar fractions in halos
where the efficiency is {\it not} suppressed, i.e.,
$\vcirc(\zsat)>\vcrit$.
We will frequently refer to the quantity
$(M_*/\Msat) \times (\Omega_m/\Omega_b)$ as the ``star formation efficiency,''
by which we mean the efficiency with which the halo converted the baryons
available to it at $\zsat$ (for a universal baryon fraction) into
stars observable at $z=0$.

\begin{figure*}
\includegraphics[width=\textwidth]{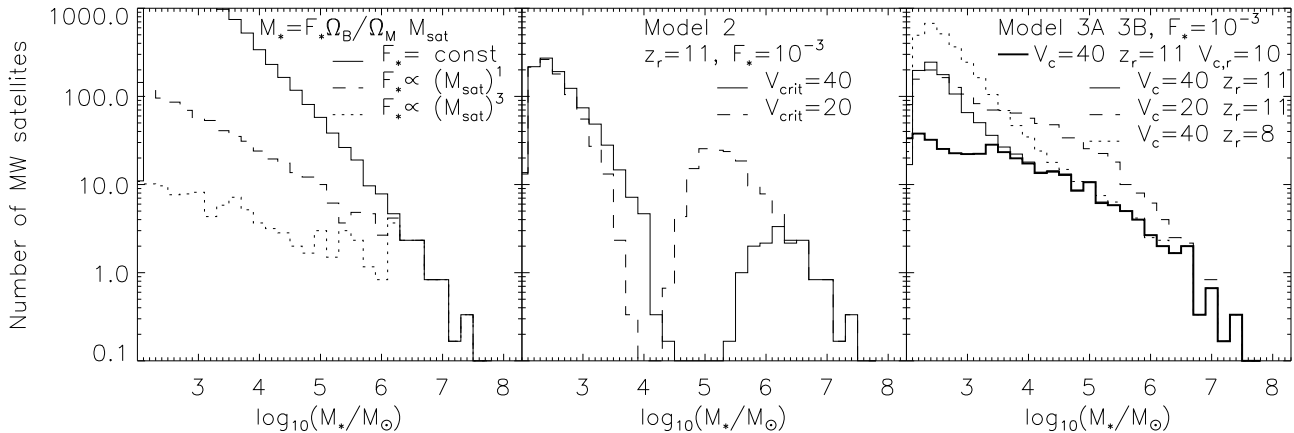}
\caption{ 
Predicted stellar mass functions of all satellites within the
MW's virial radius (280 kpc), for a variety of models.
{\it Left panel}: The solid, dotted, and
dashed lines represent, respectively,
Model~1A with $F_*=10^{-3}$ and Model~1B
with $(F_*,M_0,\alpha)=(10^{-3},10^{10}\msun,1)$ and $(10^{-3},10^{10}\msun,2)$.
{\it Middle panel}: The 
two curves show predictions of Model~2, with
$F_*=10^{-3}$, $\zrei=11$, and $\vcrit=40\kms$ ({\it solid}), and
$\vcrit=20\kms$ ({\it dashed}). 
{\it Right panel}: Thin solid, dashed, and
dotted lines represent Model~3A with
$(F_*,\vcrit,\zrei)=(10^{-3},40\kms,11)$,
$(10^{-3},30\kms,11)$, and $(10^{-3},40\kms,8)$, respectively. 
The thick solid curve shows
model~3B with $F_*=10^{-3}$, $\vcrit=40\kms$,
$\zrei=11$, and $\vcritr=10\kms$. All curves reflect the average of six
realizations of MW halos. These are the predicted {\it complete} 
satellite (stellar) mass functions, with no radial or sky coverage
selection effects.
}
\label{fig:smass_distr}
\end{figure*}

\begin{deluxetable}{cc}
\tablecaption{List of models used\label{tab:models}}

\tablehead{ \colhead{Model Name} &
\colhead{Present-Epoch Stellar Mass} }
\startdata
1A & $M_*=f_* \times \msat $\\
&\\

1B & $M_*=f_* \times {\tt min}((\msat/M_0)^\alpha,1) \times \msat $\\
&\\

2 & $
M_*=
\begin{cases}
    f_*\times \msat & \text{if }\vcirc(\zsat)>\vcrit \\
    f_*\times \mrei & \text{if
}\vcirc(\zsat)<\vcrit
\end{cases}
$ \\
&\\

3A & $M_* = \frac{f_* \times (\msat-\mrei)}{(1+0.26\,(\vcrit/\vcirc)^3)^3} +
f_*\times
\mrei$\\
 &  \\
&\\
&same as 3A for halos with $\vcirc(\zrei)>\vcritr$,
 \\3B & for halos with $\vcirc(\zrei)<\vcritr$ \\ &
$M_* = \frac{f_*
\times \msat}{(1+0.26\,(\vcrit/\vcirc)^3)^3}$ \\
\enddata
\end{deluxetable}

\subsection{Detectability and observable properties for the simulated
satellites}\label{sec:disp_modeling}

%DW: I don't think we need this first paragraph; I have omitted most
% of the paragraphs through the paper whose content is "here is what
% we just did and here is what we are going to do next."  I don't
% strongly object to including these, but I think they make the
% paper longer and more repetitive than it needs to be, with at most
% minimal gain in clarity.

%With several realizations of DM sub-halo populations in place
%(Section~\ref{sec:dm_halos}), and with set of
%possible prescriptions how to assign present-day stellar masses to 
%each of them(Section~\ref{sec:recipes}), we are now in a position to ask
%which subset of them
%at what Galactocentric distance could be found given SDSS's detection 
%efficiency\citep{lfpaper} and what the expected distribution of stellar velocity
%dispersions
%would be for those. This can then be compared to the actually observed
%distribution
%of satellite luminosities, to the distribution of their Galactocentric distances
%and
%their stellar velocity dispersions, $\sigma_*$.

Color-magnitude diagrams for the faint dwarf spheroidal galaxies in the 
Milky Way halo show that the stellar populations are predominantly `old'
(older than several Gyrs) and metal poor ([Fe/H]$\lesssim -1 $).
To convert stellar masses to luminosities, we assume that all of our
model dwarfs have a stellar mass-to-light ratio 
$M_*/L_V \approx 1 M_\odot/L_\odot$ appropriate to an old, metal poor
population \citep{bruzual03,martin08}. 
The light of the lowest luminosity dwarfs can be 
dominated by a handful of bright stars and thus subject
to stochastic variations.  We ignore this complication; our ``luminosities''
are simply scaled stellar masses: $L_{\rm V}/L_\odot = M_*/M_\odot$. This 
seems appropriate, since the luminosities of the dwarfs galaxies
are usually measured either by integrating
over the luminosity function of old stellar population matched to the
observed luminosity function of stars in dwarfs\citep{belokurov06}
or by averaging over possible stochastic variations of galaxy 
luminosity\citep{martin08}.

The detectability of a faint stellar MW satellite galaxy in an SDSS-like search
depends on its luminosity and its distance from the Sun, as quantified by
\citet{lfpaper} \citep[see also][]{walsh08}. On the basis
of these results (Figure 12 of \citealt{lfpaper}) we model the detectability of
each 
simulated satellite as a binary decision using the criterion
\begin{equation}
{\rm log}_{10}(D_\sun/1\,{\rm kpc})<1.1-0.228 M_V 
\end{equation}
Our simulations provide
the current Galactocentric distance and orbital
apocenter and pericenter for each sub-halo, but not the
orientation of the orbit.
We therefore assign the heliocentric distance of 
the satellites 
\begin{equation}
D_\sun=\sqrt{8.5^2+D_{GC}^2-2\times8.5\times D_{GC}\,{\rm cos}(\phi)} ~,
\end{equation}
where
$D_{GC}$ is the Galactocentric distance (in kpc)
from the simulations and ${\rm cos}(\phi)$ is
a random variable uniformly distributed between $-1$ and $1$ ($\phi$ is the
angle
between radial vectors from the GC to the Sun and to the sub-halo).
This method
assumes that the satellite orbits are isotropically distributed across the
sky \citep[see][for discussion of the validity of this 
approximation]{tollerud08}.
As expected
from \citet{lfpaper},
accounting for the detectability of satellites causes the `observable'
population to differ strongly from the `simulated' one; only the brightest
satellites are observable throughout the virial volume. 

Not surprisingly, the \citet{lfpaper} analysis also reveals a surface
brightness threshold for dwarf detection, which is approximately
30 mag arcsec$^{-2}$ with little dependence on distance.
We assume that any model dwarf that passes the luminosity threshold
also passes the surface brightness threshold. 
Many recent SDSS satellite discoveries do lie near that survey's surface
brightness limit; this assumption can therefore only be tested with the
next generation of sky surveys.  We discuss implications of this
assumption in \S\ref{sec:conclusions}.

With a model that assigns stellar luminosities to each satellite halo, we can 
predict the expected stellar velocity dispersions for comparison with 
those measured for MW satellites by \citet{walker07}, \citet{simon07},
and \citet{martin07}.
This can be done straightforwardly if we assume that the stars are 
test particles --- an assumption supported by the observed 
$(M/L)_{\rm dyn}(<R_{\rm eff}) \gg (M/L)_*(<R_{\rm eff})$ ---
orbiting in an NFW potential
with an isotropic velocity dispersion. Then we can use the Jeans
equation \citep{jeans19} to derive the velocity dispersion profile of stars:
\begin{equation}
\frac{d(\nu(r)\,\sigma^2(r))}{d\,r} + \nu(r) \, \frac{G\,M(r)}{r^2}=0,
\end{equation}
where $\nu$ is the density distribution of stars 
(see \citealt{strigari07} for more detailed treatment).
Here we assume that the
density of stars follows a Plummer profile
$\nu \propto [1+(r/r_p)^2]^{-2}$ \citep{plummer11}, which 
seems to fit observed density profiles reasonably well
\citep{wilkinson02,belokurov07}. The mass profile $M(r)$ used here is computed 
based on the virial radii and concentrations at the redshift $z_{\rm sat}$
of sub-halo accretion.
While the outer parts of the sub-halos are tidally stripped,
\citet{penarrubia08} show that the stars and the 
inner part of the dark matter sub-halo are
stripped only at a very late stage, when the sub-halo
is close to complete disruption.
They also show that the velocity dispersion 
in sub-halos is a function of the total dark matter mass 
remaining bound inside the
luminous body and therefore remains nearly constant until this
late stage.

After numerically solving  the Jeans Equation,
we compute the expected light-weighted velocity dispersion within the optical
radius as
\begin{equation}
\sigma_*=\frac{\int \nu(r)\sigma(r) \,dx\,dy\,dz}{\int \nu(r) \,dx \,dy \,dz},
\end{equation}
where the
integration is done over a cylinder within a
radius, $R = \sqrt{x^2+y^2}$ equal to the Plummer
radius of the galaxy; the integral extends over $\pm \infty$ in $z$.
The stellar velocity dispersion depends 
on the radial extent of the stellar tracers, which
cannot be predicted within
our simple modeling context \citep[see also][]{benson02}. We therefore
use the {\it observed} properties of the faint Milky Way satellites
to choose stellar radii, based on \citet{martin08}.
Specifically, we adopt Plummer radii
$r_p = 150$ pc for $M_V<-5$, and for fainter dwarfs we adopt
a linear relation between log $r_p$ and $M_V$ with
$r_p$ rising from 20 pc at $M_V=0$ to 150 pc at $M_V=-5$.

The additional important component of the detectability is the tidal disruption
of the satellite galaxies. Although our semi-analytic model of dark matter
sub-halo evolution properly accounts for the tidal disruption of sub-halos,
it does not allow for the possibility that stars have been dispersed
in a tidal stream while a small core of the sub-halo survives.
Here we simply classify a sub-halo as unobservable if its current tidal
radius is less than the expected
Plummer radius of the stellar body, which 
would imply substantial tidal disruption of the stellar component.
We also presume that a satellite is unobservable if its host sub-halo
has lost more than 99\% of its original mass to tidal stripping.

\section{Results}\label{sec:results}
 
%In \S~\ref{sec:populating_halos} we have laid out our approach to
%combining semi-analytics models for the DM sub-halo population with 
%established ``recipes" to ascribe (or suppress) the formation
%of stars within them. Now we proceed to work out the observable
% consequences, starting with the predictions for the complete
%satellite population in \S~\ref{sec:stellar_mass_distr}, then
%focussing on the models predictions for the number of satellites
%detectable with the current generation of sky surveys, Nobs . 
%In \S~\ref{sec:observed_distribution} we move beyond nbs , comparing
%also the distributions of stellar velocity dispersion and 
%of satellite distances. 

\subsection{Stellar mass function of the full satellite
populations}\label{sec:stellar_mass_distr}
%We start the analysis  of the satellite and DM halo discrepancy by illustrating
%how the different prescriptions for the
%star-formation histories of low-mass halos from Section~\ref{sec:recipes}
%translate into a predicted stellar mass function of Milky Way satellites,
%averaged over the six Monte-Carlo realizations.

Figure~\ref{fig:smass_distr} shows the 
predicted distribution of the {\it stellar} masses of satellites  within
R$_{\rm virial}= 280$ kpc, assuming 4$\pi$ sky coverage and complete satellite
detectability. In the left panel, the solid curve shows Model 1A with
a constant $F_* = 10^{-3}$, making the stellar mass function a scaled
version of the dark matter sub-halo mass function.  Introducing 
mass-dependent suppression, Model 1B with $\alpha=1$ (dashed) and
$\alpha=2$ (dotted) lowers the low mass end of the stellar mass
function as expected.  Since this model
also adopts $F_*=10^{-3}={\rm const.}$ above $\msat=M_0=10^{10} M_\odot$,
the high mass end of the mass function is unchanged.

The middle panel of Figure~\ref{fig:smass_distr} shows Model 2, with
post-reionization suppression of star formation in halos below
a sharp circular velocity threshold, either 
$\vcrit=40\kms$ (solid) or $\vcrit=20\kms$ (dashed), where we have
adopted $F_* = 10^{-3}$ and a reionization redshift $\zrei=11$.
The resulting stellar mass functions for the satellite galaxies
are strongly bimodal, with the low mass
portion corresponding to dwarfs in which all stars formed before
reionization and the high mass portion corresponding to halos
that exceeded the critical velocity threshold before becoming satellites,
$\vcirc(\zsat)>\vcrit$.  The low mass portion is just a scaled version
of the sub-halo mass function at $z=\zrei$.
Above $M_* \approx 10^{6.5}M_\odot$ the host halos are all massive
enough to have star formation after $\zrei$, and the mass function
is the same as
that of Model 1.  If the velocity threshold is lowered to
$\vcrit=20\kms$, the high mass peak in the distribution of
satellite stellar masses extends to lower values
before photo-ionization suppression cuts it off.

The bimodal appearance of the middle panel of
Figure~\ref{fig:smass_distr} is a direct consequence
of the sharp $\vcirc$ threshold for photo-ionization suppression.
The right hand panel shows predictions for
several variants of Model 3A and 3B, with the 
\cite{gnedin00} formula (Eq.~\ref{eq:gnedin_recipe_vc})
used to describe photo-ionization suppression.
With this smooth suppression, the ``pre-reionization'' and
``post-reionization'' portions of the mass function join to
make a smooth overall mass function.
The low mass end of the mass function is now a mix of satellites
that formed their stars before reionization and satellites
with $\vcirc(\zsat) < \vcrit$ whose post-reionization 
star formation was strongly suppressed but not completely eliminated.
Lowering the assumed
reionization redshift from $\zrei=11$ to $\zrei=8$
boosts the stellar mass function below $M_*=10^4 M_\odot$.
Conversely, if we eliminate pre-reionization SF in dwarfs with
$\vcirc(\zrei) < \vcritr = 10\kms$ (thick solid line, Model 3B), the number
of satellites with $M_* \leq 10^3 M_\odot$ drops by a large factor,
while at higher masses the stellar mass function is unaffected.
The difference between the thin and thick solid lines
is the contribution of satellites that formed
stars primarily before reionization in halos with $\vcirc(\zrei ) < 10\kms$,
for $\zrei=11$ and $\vcrit=40\kms$.

%Much of the impact that these various recipes have on the distribution
%of stellar masses for the Milky Way have already been discussed by other
%authors. We now move beyond this by incorporating the satellite 
%detectability, allowing a more rigorous comparison of models with the
%observed satellites.

\subsection{Distribution of observed dwarf satellite luminosities, $N(M_V)$}
\label{sec:observed_distribution}

\begin{figure}
\includegraphics[width=0.5\textwidth]{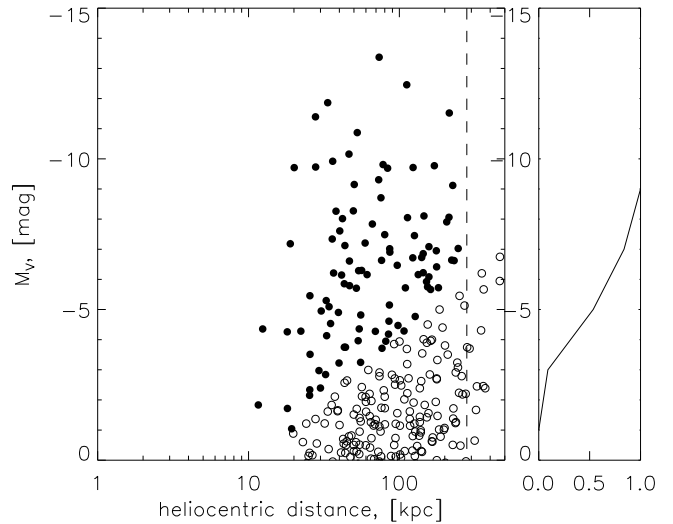} 
\caption{Detectability of the satellite galaxies predicted by our fiducial
model (Model 3B), as a function of their heliocentric distance and
stellar luminosity.  Filled circles denote galaxies that
can be detected with SDSS-like all-sky surveys, and
empty circles denote those that cannot. The dashed line marks the 
approximate virial radius of the MW's dark matter halo;
we will compare all model predictions to the 
observed MW satellite population only within this radius. 
The galaxies shown were taken
from one Monte-Carlo realization of Model~3B with
($\vcrit,F_*,\zrei,\vcritr)=(35\kms,10^{-3}, 11, 10\kms)$. The right
panel shows the
fraction of detectable galaxies as a function of luminosity.}
\label{fig:mv_vs_dist}
\end{figure}

Figure~\ref{fig:mv_vs_dist} illustrates the impact of 
selection effects on the observable satellite population.
For one realization of Model 3B (with parameters that yield
a good match to observations), filled circles show satellites
that would be detectable in an all-sky, SDSS-like survey \citep{lfpaper},
and open circles show un-detectable satellites.
The low end of the luminosity distribution, with $M_V \ga -5$,
is strongly affected by the radial selection bias.

\begin{figure*}
\includegraphics[width=0.5\textwidth]{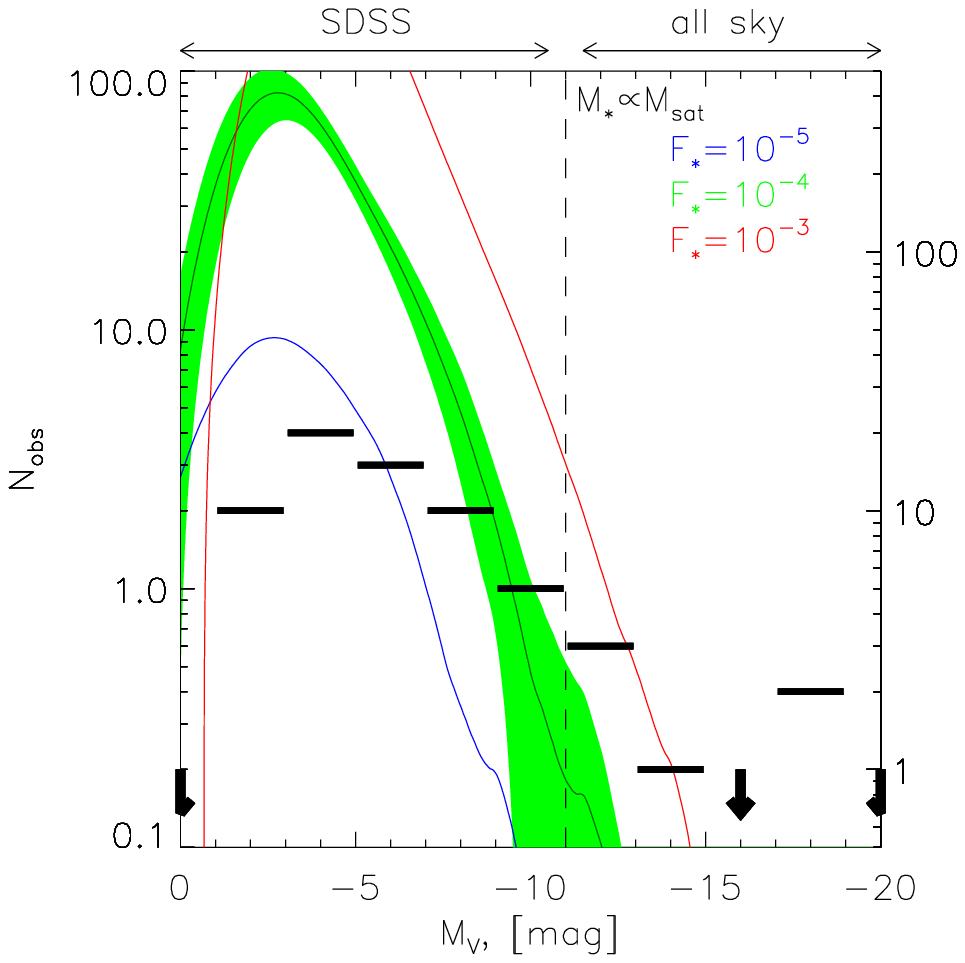}
\includegraphics[width=0.5\textwidth]{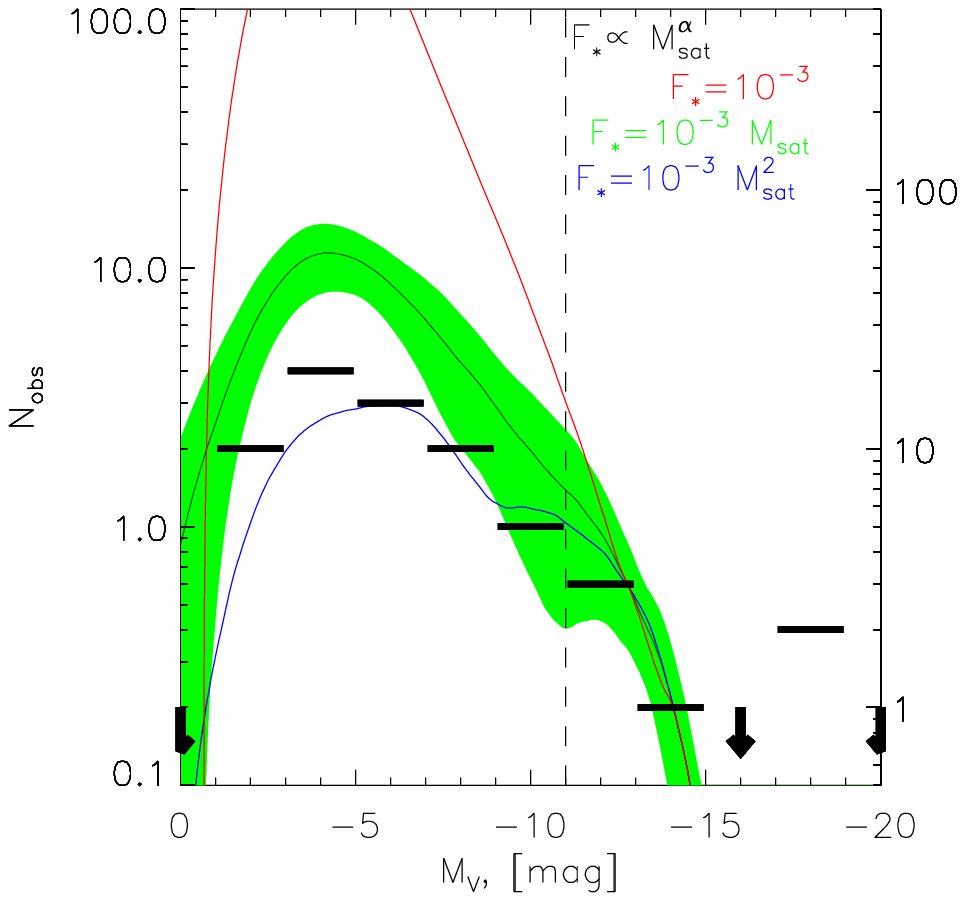}
\caption{Model predictions for the observed satellite
population, $\Nobs$, including radial selection effects for the
SDSS dwarfs.
Horizontal bars show the number of currently known satellites
(Table 2) in 2-magnitude bins; empty bins are plotted with an arrow.
The SDSS and classical dwarfs are separated by the vertical line
at $M_V=-11$; note that the $y$-axes for these two populations
differ by a factor of five so that the model predictions (which
incorporate a factor of 1/5 below $M_V=-11$ to account for SDSS
sky coverage) are continuous across the boundary.
{\it Left Panel:} Predictions of Model 1A, with $M_* \propto \Msat$,
for three values of $F_*$.  For $F_* = 10^{-4}$, the green band shows
the bin-by-bin $\pm 1\sigma$ range of the predictions from multiple 
realizations; the logarithmic width of this band is similar for
other models.  Model curves have been slightly smoothed with a 
polynomial filter.
{\it Right Panel:} Comparison of Model 1A (red curve) to Model 1B,
where the stellar mass fraction in halos with $\Msat < 10^{10} M_\odot$ is
is $F_* \propto \Msat^\alpha$, with
$\alpha=1$ (green band) or $\alpha=2$ (blue curve).
}
\label{fig:hypo1_mv}
\end{figure*}

\begin{figure}
\includegraphics[width=0.5\textwidth]{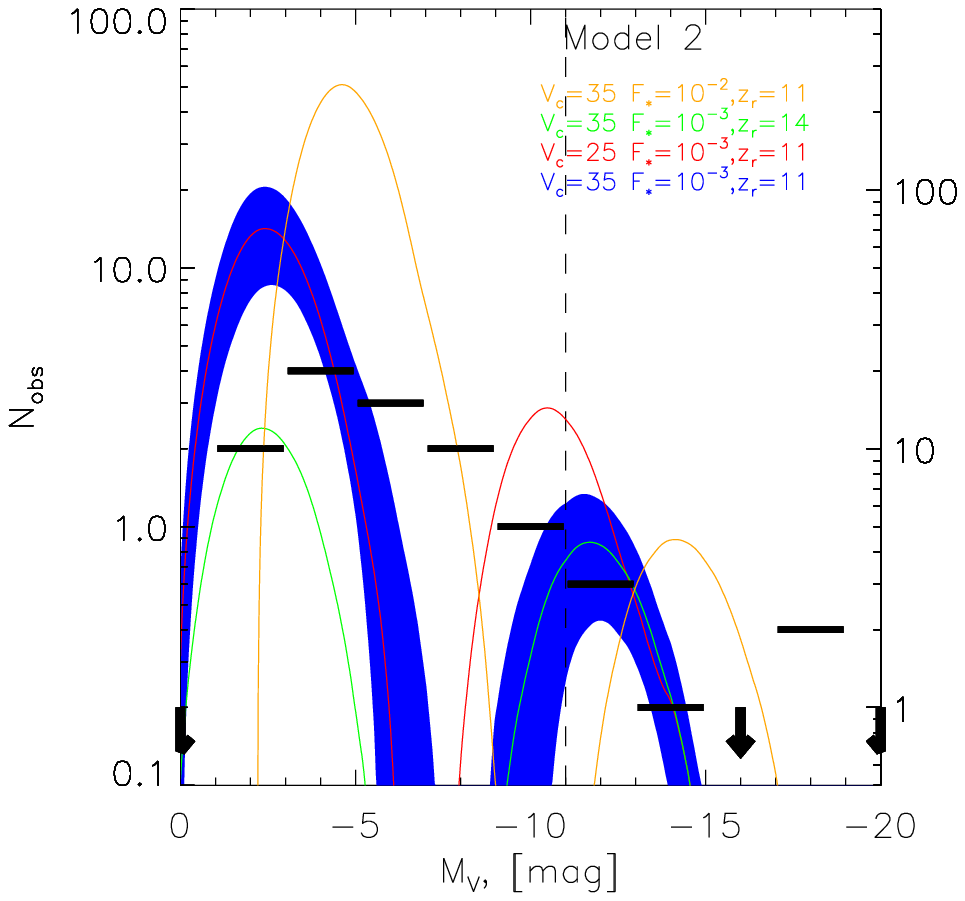}
\caption{Predictions for Model 2, in which post-reionization star
formation is sharply suppressed below a critical velocity $\vcrit$,
in the same format as Figure~\ref{fig:hypo1_mv}.
Blue, red, green, and orange curves/bands show the parameter combinations
($F_*, \vcrit, \zrei)=(10^{-3}, 35\kms, 11$),
($10^{-3}, 25\kms, 11$),
($10^{-3}, 35\kms, 14$), and
($10^{-2}, 35\kms, 11$), respectively.
This class of models predicts a bimodal distribution of satellite 
luminosities, with the faint portion ($M_V>-8$) coming entirely
from pre-reionization star formation. The predicted $N(M_{\rm V})$ differs
grossly from the observations.}
\label{fig:sharpv_mv}
\end{figure}

\begin{figure}
\includegraphics[width=0.5\textwidth]{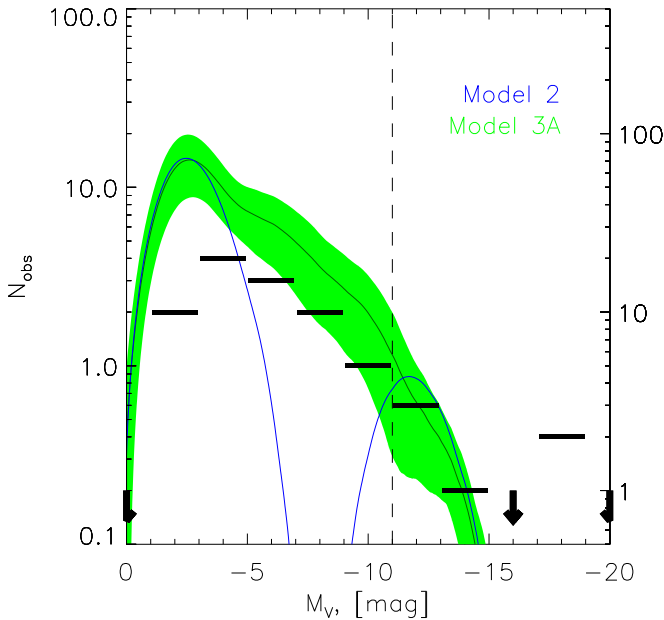}
\caption{Comparison of Model~2 and Model~3A, both with
parameters $F_*=10^{-3}$, $\vcrit=35\kms$, and $\zrei$=11,
in the same format as Figure~\ref{fig:hypo1_mv}.
Switching to the continuous prescription for photo-ionization
suppression fills in the gap between the two peaks of Model~2,
while leaving the predictions at the highest and lowest
luminosities unchanged.
}
\label{fig:sharpv_gnedin_mv}
\end{figure}

\begin{figure*}
\includegraphics[width=0.5\textwidth]{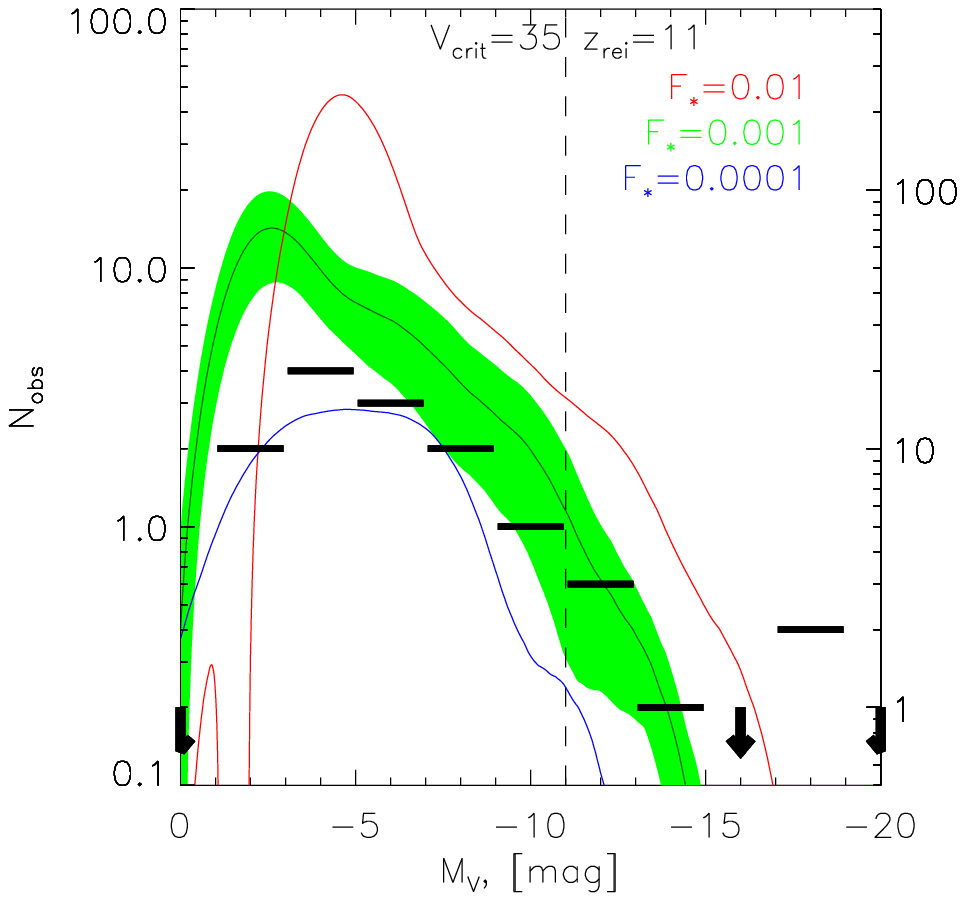}
\includegraphics[width=0.5\textwidth]{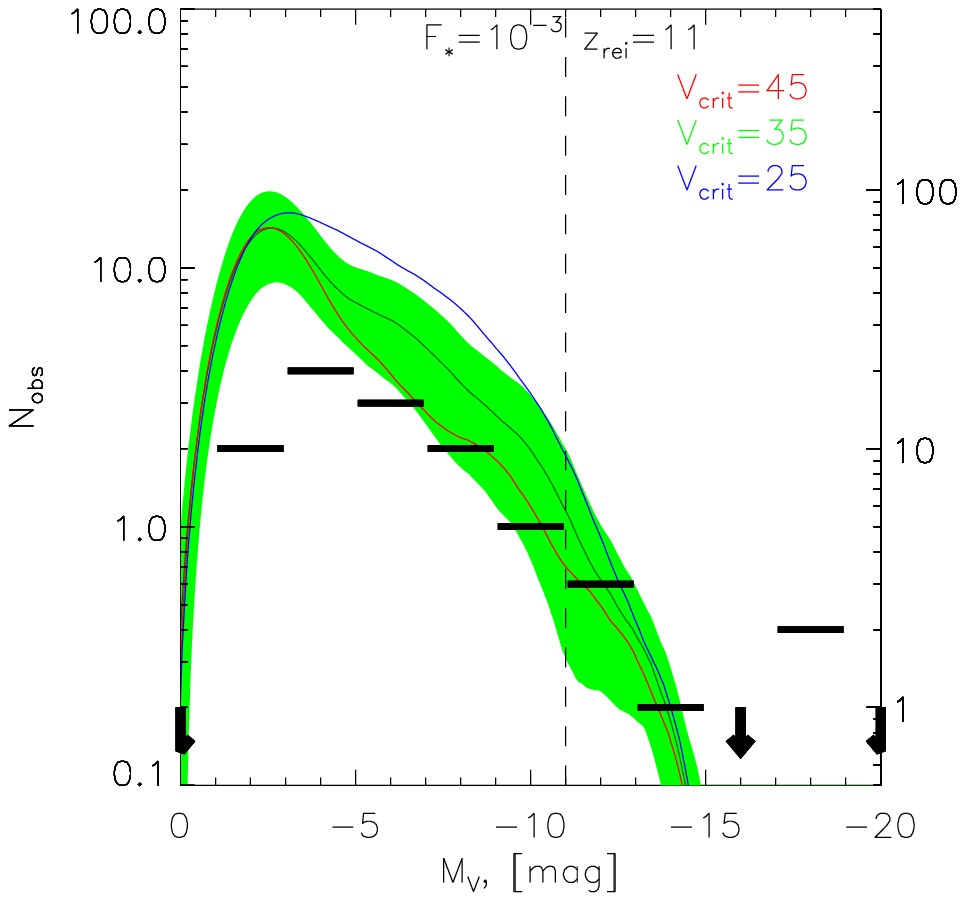}
\includegraphics[width=0.5\textwidth]{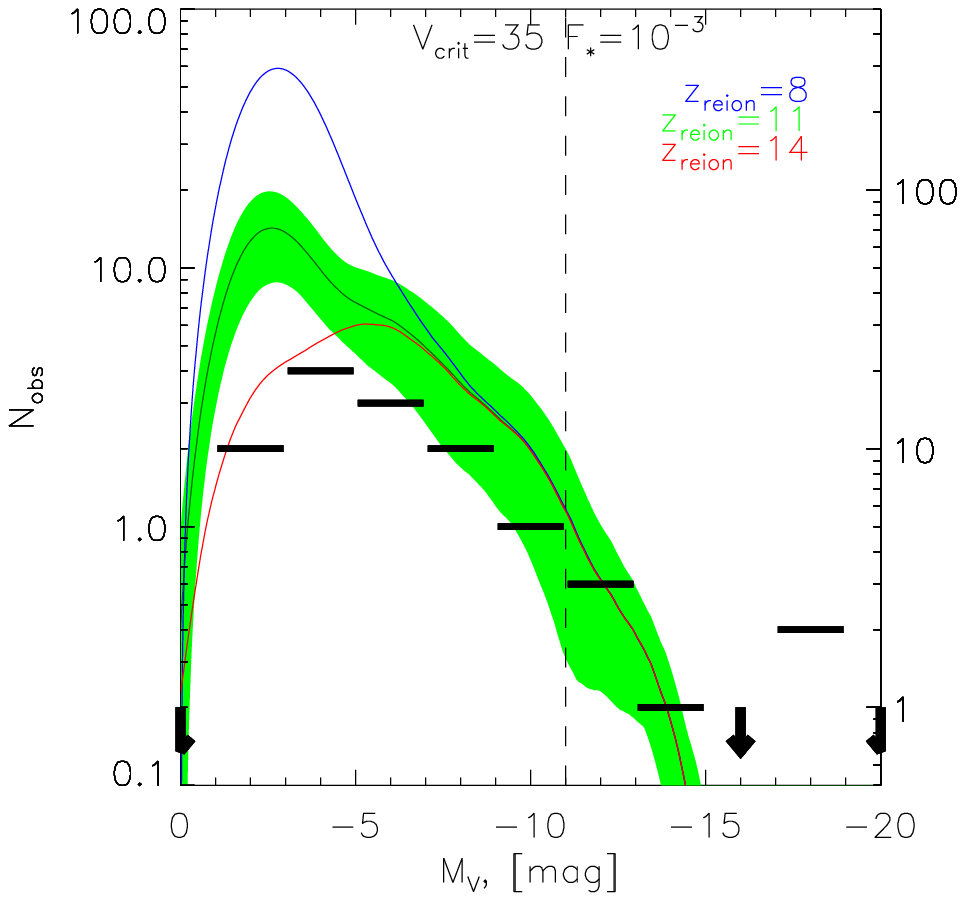}
\includegraphics[width=0.5\textwidth]{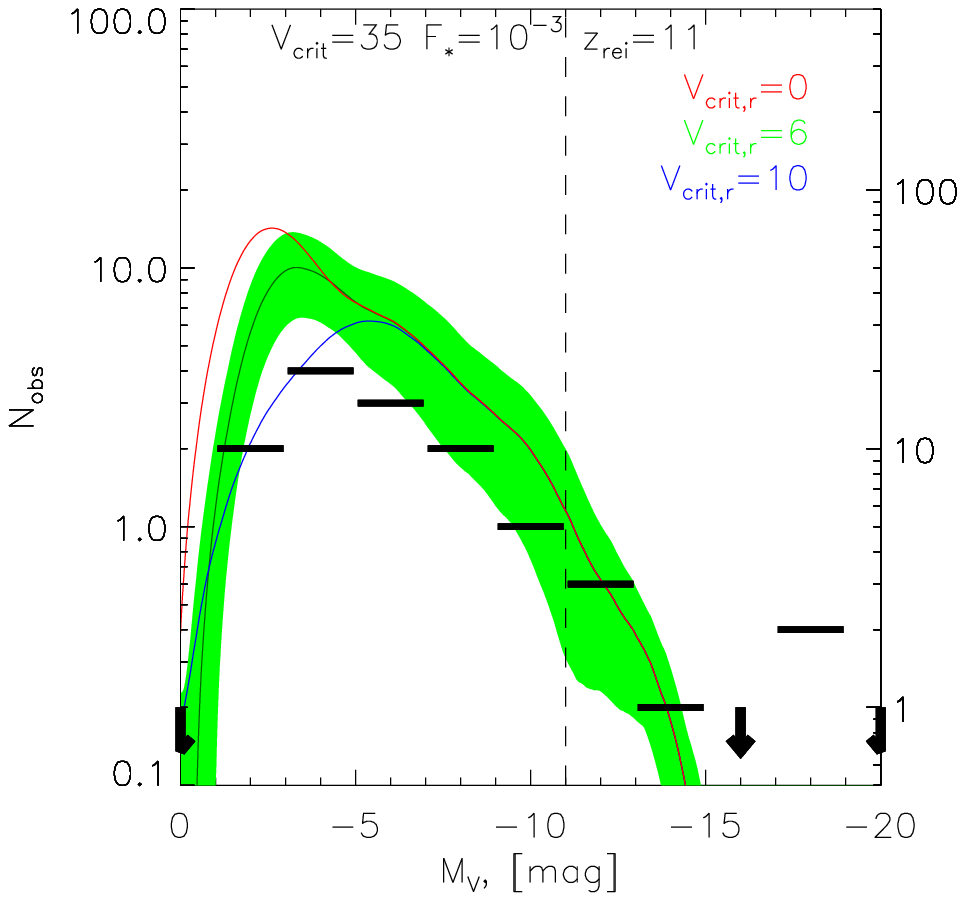}
\caption{
Predicted $\Nobs$ for Models~3A and~3B with a variety of parameter choices,
in the same format as Figure~\ref{fig:hypo1_mv}.
In the first three panels, green bands show Model~3A predictions for a
reference parameter set 
$F_*=10^{-3}$, $\vcrit=35\kms$, $\zrei=11$.
Red and blue curves show the impact of changing
the stellar mass fraction to $F_* = 10^{-2}$ or $10^{-4}$
(top left), the critical velocity threshold to 
$\vcrit = 45\kms$ or $25\kms$ (top right), or the
reionization redshift to $\zrei=8$ or 14 (lower left).
The lower right panel compares the prediction of this reference
model (now shown by the red curve) to predictions of Model~3B
with a pre-reionization critical threshold 
$\vcritr = 6\kms$ (green band) or $10\kms$ (blue curve).
}
\label{fig:gnedin_mv}
\end{figure*}

\begin{figure}
\includegraphics[width=0.5\textwidth]{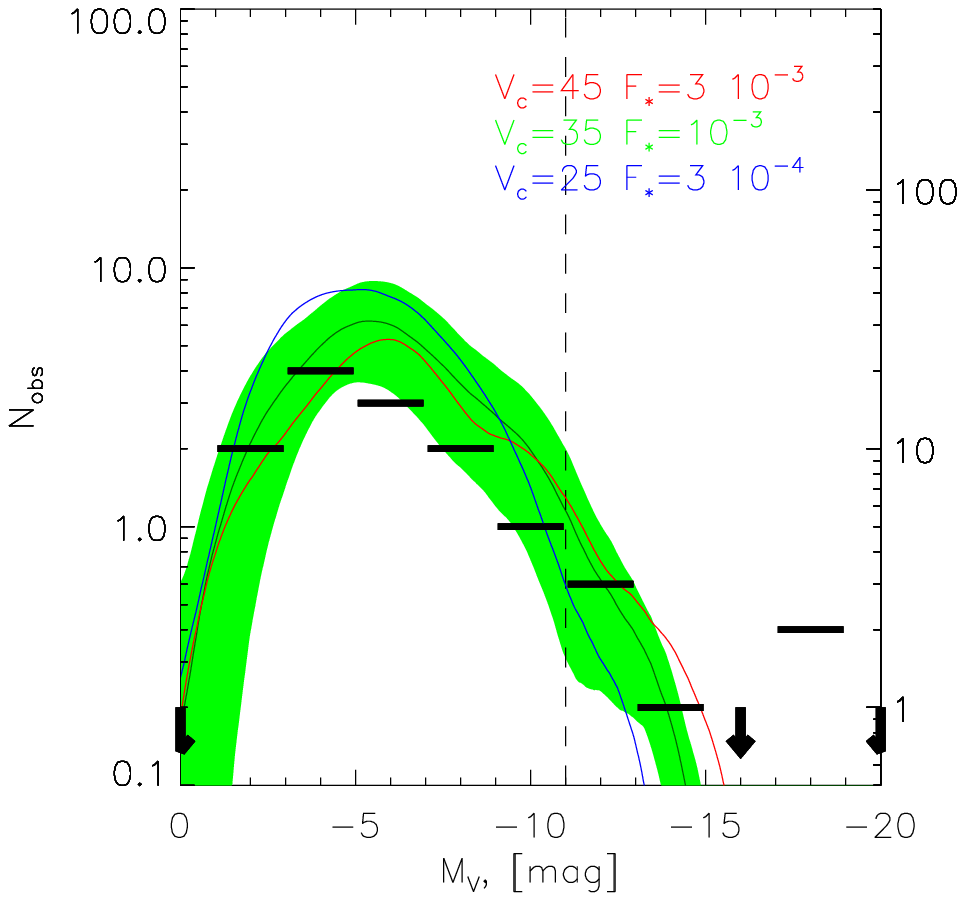}
\caption{Degeneracy between $F_*$ and $\vcrit$ for Model 3B,
in the format of Figure~\ref{fig:hypo1_mv}.
Blue, green, and red curves/bands show the parameter combinations
($F_*,\vcrit)=(3\times 10^{-4}, 25\kms)$,
$(10^{-3},35 \kms)$, and $(3\times10^{-3},45 \kms)$, 
which all yield similar levels of agreement with the observations.
We adopt $\zrei=11$ and $\vcritr=10\kms$ in all cases.
}
\label{fig:gnedin_combinations}
\end{figure}

For direct comparison with observations, we therefore select only
those model satellites whose combination of luminosity and distance
would make them detectable.
At the bright end, $M_V < -11$, we assume that 
existing photographic surveys are complete to $D_\odot = 280\,$ kpc, and 
we thus compare the total number of dwarfs
across the whole sky to the total population of satellites within the 
virial radius in the simulation.
For $M_V \geq -11$, we randomly select 1/5 of the model galaxies
to mimic the 20\% sky coverage of SDSS DR5, and we count only
those satellites that would be detectable according to the criteria
of \cite{lfpaper}.
We focus our data-model comparison on $\Nobs$, the luminosity distribution
of known MW satellites.
We look at additional tests against 
stellar velocity dispersions, central masses, and the
heliocentric radial distribution in 
\S~\ref{sec:veldisp_radial}.

The luminosities, distances, and velocity dispersions of the 
observed  Milky Way
satellites that we use in all subsequent model - data comparisons
were taken from
various authors~\citep{mateo98,metz07,martin08} and are compiled in
Table~\ref{tab:satellite_list}. The sample of SDSS satellites used here
consists of those systems above the 50\% completeness limits of~\citet{lfpaper}. 
We do not include two systems, BooII and LeoV \citep{walsh07,belokurov08}, which
do not formally satisfy the very conservative selection limits from \citet{lfpaper}
These limits were chosen to avoid the issue of significant 'false positive' 
detections, at the expense of leaving out 2 objects that deeper follow-up 
found to be 'real'. For the analysis presented here it is most important
that the {\it same} selection criteria are applied to the mock satellite
observations and the SDSS data. As our analysis subsequently shows, such
a small difference in sample size is smaller than the model halo to
halo variation of number of galaxies. Therefore the inclusion of omission
of these two objects  does not affect our results significantly.

 Anyway, as we will see later, the halo to halo
variation of number of galaxies in our models is noticeable, so we
believe that the fact that we do not include two 
galaxies should not affect our results significantly. 

The left panel of Figure~\ref{fig:hypo1_mv} compares our simplest model
($M_* \propto \Msat$, Model 1A) to the observed satellite counts,
now including the satellite galaxy selection effects in the model.
We randomly sample each of the six Monte Carlo halo simulations 
five times (choosing 1/5 of the faint satellites but
always keeping the full set for $M_V < -11$), 
compute the mean model prediction as the mean of these 30 samplings,
and compute the rms dispersion among these 30 in each absolute
magnitude bin.
%{\bf DW: Is this an accurate description?  Jaiyul's right that it
%isn't actually the correct thing to do, though perhaps the difference is
%small enough that we don't need to worry about it.
%The main problem is that this procedure will underestimate
%the standard deviation for bright satellites by a factor of sqrt(5);
%it also fails for faint satellites if the halo-to-halo variations
%are significantly different from Poisson.  The right thing to do is
%compute the mean from all satellites (with a division by five instead
%of random sampling for
%the faint ones) and compute the dispersion from one random sampling
%of each of the six halo realizations.}
Despite the selection bias against low luminosity satellites, this
model fails drastically for any choice of $F_*$, predicting a much
steeper luminosity function than observed.
For example, the model with $F_* = 10^{-4}$ matches the observed
counts near $M_V=-9$ but predicts far too many satellites fainter
than $M_V = -6$.  Selection effects and newly
discovered satellites have
not altered this basic discrepancy, first emphasized by~\citet{klypin99}
and ~\citet{moore99}. 
The green band shows the $1\sigma$ dispersion in predicted
counts, and it is clear that statistical fluctuations will not
resolve the discrepancy either.

In the right panel we apply our purely empirical modification,
$M_*/\Msat \propto M^\alpha$ below a halo mass $\Msat=M_0=10^{10} M_\odot$
(Model 1B).  With $F_*=10^{-3}$ and $\alpha=2$, this model 
achieves reasonable agreement with the
the observed $\Nobs$ over the full
range $0 \geq M_V \geq -15$.  The agreement can be further improved
by adjusting $F_*$ and $M_0$, so it appears that this level of
mass-dependent suppression is approximately what is needed to 
explain the observed shape of $\Nobs$.  Linear suppression
($\alpha=1$, green band) is not sufficient, predicting an excess
of faint dwarfs when normalized to the bright dwarfs.
All of our models fail to match the brightest bin (comprised of
the SMC and LMC); we defer discussion of this discrepancy
to the end of this Section.

Figure~\ref{fig:sharpv_mv} shows the expected $\Nobs$
distributions for Model~2, which has a sharp $\vcrit$ threshold for 
the suppression of SF after reionization in small halos. 
As in Figure~\ref{fig:smass_distr}, the predicted $\Nobs$ is
bimodal, with a bright peak corresponding to halos that exceeded $\vcrit$
before $\zsat$ and a faint peak corresponding to stars formed
before reionization.  Raising the stellar fraction $F_*$ with
other parameters fixed (orange vs.\ blue) shifts both peaks
horizontally to higher $M_V$; the faint peak also increases in
height because the brighter (though still faint) satellites
can be seen over a larger fraction of the MW virial volume.
Lowering $\vcrit$ with other parameters fixed (red vs.\ blue)
has no impact on the faint peak, but the bright peak extends
to fainter magnitudes and grows in height because lower mass
halos can now be populated with stars after reionization.
Raising $\zrei$ (green vs.\ blue) with other parameters fixed
has no impact on the bright peak, but it shifts the faint
peak downwards in amplitude and slightly downwards in location
because halos have accreted less mass by this higher redshift.
While photo-ionization suppression reduces the discrepancy
with the number of faint satellites seen in Model~1A, these
sharp threshold models predict a gap between the faint and
bright satellites that is clearly at odds with the data.

Figure~\ref{fig:sharpv_gnedin_mv} compares the Model~2 predictions
with those of Model~3A, which uses the \cite{gnedin00} formula to
incorporate a smoothly increasing suppression of the stellar 
mass fraction in halos with $\vcirc(\zsat) \la \vcrit$.
In both cases we use parameters
$F_* = 10^{-3}$, $\vcrit=35\kms$, $\zrei=11$.
Model~3A is more physically realistic than Model~2, with a
mass-dependent suppression that is calibrated on numerical
simulations (and is approximately consistent with three
independent numerical studies).
Galaxies formed in halos with $\vcirc(\zsat) \la \vcrit$
now fill the gap that was present in Model~2, producing a
luminosity distribution that rises continuously from
$M_V = -14$ down to $M_V = -2$, before radial selection
effects finally cut it off.
With these parameter choices, pre-reionization dwarfs 
dominate the counts (and exceed the observations)
for $M_V \leq -4$, but suppressed
post-reionization dwarfs dominate the counts at all
brighter magnitudes.

Since Model~3 is both more physically realistic and more
empirically successful than Models~1 and~2, we focus on 
it for the remainder of the paper, including Model~3B in
which pre-reionization star formation is suppressed below
a circular velocity threshold.
Figure~\ref{fig:gnedin_mv} systematically explores the
impact of parameter variations in Models~3A and~3B.
In the first three panels, the green band shows the Model~3A
predictions for a fiducial set of parameter choices,
$F_* = 10^{-3}$, $\vcrit=35\kms$, and $\zrei=11$.
Changing $F_*$ (top left) shifts the predicted distribution
horizontally to higher or lower luminosities, with some
change in shape at the faint end because of the luminosity
dependence of radial selection effects.
Changing $\vcrit$ alters the predicted counts at intermediate
luminosities, $-4 > M_V > -11$, while having little effect
at the faint end (where pre-reionization dwarfs dominate)
or at the bright end (where most galaxies exceed the highest
threshold considered here).
Changing $\zrei$ alters the height of the pre-reionization
peak at faint luminosities but has minimal impact for $M_V < -7$.

With our fiducial parameter choices, Model~3A substantially
overpredicts the number of satellites with $M_V \approx -3$.
Raising the reionization redshift to $\zrei=14$ erases this
discrepancy, but this value of $\zrei$ seems implausible
given the strong and rapidly evolving opacity of the
intergalactic medium at $z\approx 6$ seen in quasar spectra
\citep{fan06}, and it is only marginally consistent with
the WMAP5 results.  In the lower right panel, we return
to $\zrei=11$ but suppress pre-reionization star formation
in halos with $\vcirc(\zrei) < 6\kms$ (green) or $10\kms$ (blue),
motivated by the inefficient gas cooling expected below the
threshold for atomic line excitation (Model 3B).
The $\vcritr=10\kms$ model yields acceptable agreement 
with the observed number counts over the full range
$0 \geq M_V \geq -15$.  
The $\vcritr=6\kms$ model still yields an excess of faint
satellites; results for $\vcritr=8\kms$ (not shown) are 
nearly identical to those for $10\kms$, 
indicating that an $8\kms$
threshold is already sufficient to essentially eliminate the
contribution of pre-reionization dwarfs.
This pre-reionization suppression appears to be critical 
to explaining the number of dwarfs observed by the SDSS.

Within Model~3B, there is strong degeneracy between the
values of $F_*$ and $\vcrit$.  
Figure~\ref{fig:gnedin_combinations} shows that 
the parameter combinations 
$(F_*, \vcrit) = (3\times 10^{-3}, 45\kms)$,
$(10^{-3}, 35\kms)$, and
$(3\times 10^{-4}, 25\kms)$ all yield
similar predictions and acceptable agreement with the 
observed number counts.
The lower values of $\vcrit$ are favored by the
numerical studies of \citet{hoeft06} and \citet{okamoto08}.
For the remainder of the paper we will adopt
($F_*, \vcrit, \zrei, \vcritr)=( 10^{-3}, 35\kms, 11, 10\kms)$  
as the fiducial parameter values for Model 3B.

\begin{figure}
\includegraphics[width=0.5\textwidth]{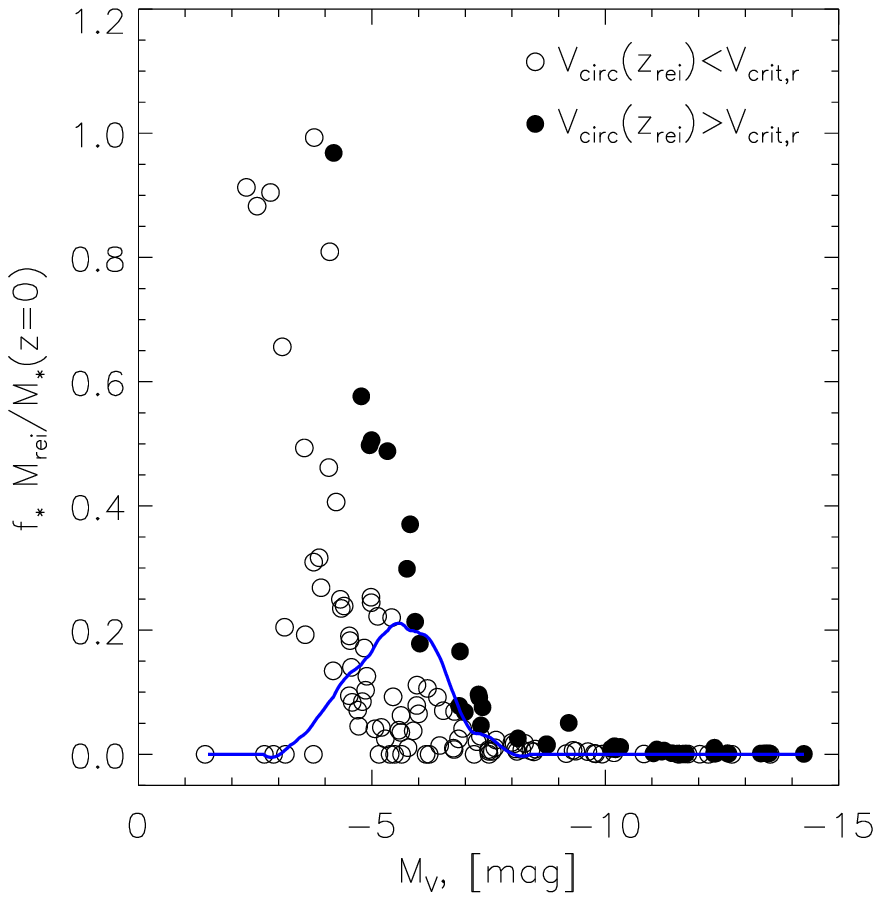}
\caption{
Fraction of pre-reionization stars in observable
satellites of different luminosities, as predicted by the fiducial
Model~3B.  Filled circles show $f_* M(\zrei) / M_*(z=0)$, the
fraction of the stellar mass that formed by $\zrei$, for 
systems that exceeded the pre-reionization threshold,
$\vcirc(\zrei)>\vcritr=10 \kms$.  
Open circles show $f_* M(\zrei) / M_*(z=0)$ for systems
with $\vcirc(\zrei) < \vcritr$, but in the context of
Model~3B these systems do not form any stars before reionization.
The curve shows the fraction of satellites that formed more
than 10\% of their stars before the epoch of reionization,
in bins of luminosity.
}
\label{fig:reion_masses}
\end{figure}

\begin{figure}
\includegraphics[width=0.5\textwidth]{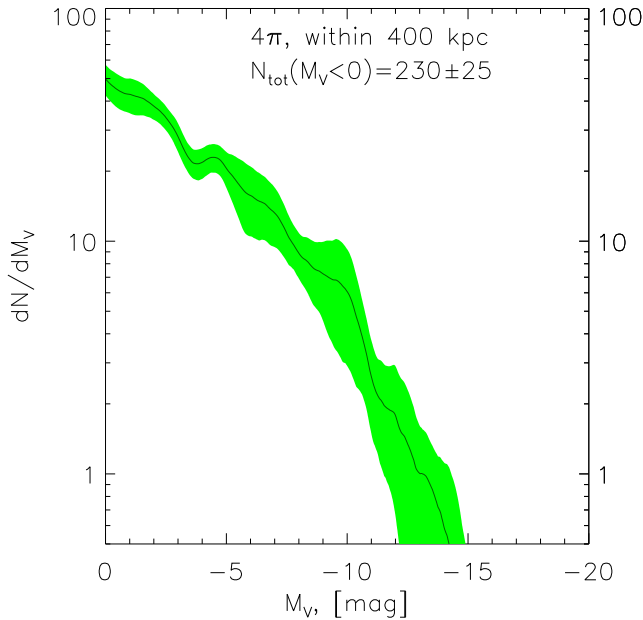}
\caption{The predicted number of MW satellites per unit magnitude within 400
kpc across the whole sky averaged from 6 MC realizations, using the fiducial 
model parameters (Model 3B with $F_*= 10^{-3}$, $\vcrit = 35\kms$,
$\vcritr = 10\kms$, and $\zrei=11$) and assuming no observational 
incompleteness.  The
total number of satellites with stellar luminosities brighter
than $M_V=0$ is $230\pm 25$.
Note that this Figure gives counts in 1-magnitude bins rather than
the 2-magnitude bins used in earlier Figures.
}
\label{fig:lf_nosel}
\end{figure}

For this fiducial model, Figure~\ref{fig:reion_masses} illustrates
in more detail the relative importance of stars formed before
and after reionization.  For systems with $\vcirc(\zrei)>\vcritr$,
filled circles show the fraction of their stars that formed before
reionization.  For systems with $\vcirc(\zrei)<\vcritr$, open circles
show the fraction of stars that {\it would} have formed before
reionization, but because of the $\vcritr$ threshold these 
galaxies have no pre-reionization stars in this model.
At every satellite luminosity, 
the average fraction of pre-reionization stars is small, or even zero, but
albeit for different reasons at high and low luminosities.
The host halos for the brighter, ``classical'' dwarf
satellites were typically massive enough at $\zrei$
to exceed $\vcrit$, but that initial population of stars
was subsequently swamped by the much larger post-reionization
population. In contrast, the halos that now host
the very faintest known satellites ($M_V>-4$)
did not exceed $\vcritr$ at $\zrei$ and hence --- in Model~3B ---
did not form any stars before $\zrei$. 
A small fraction of the satellites with $M_V \approx -5$ have
large populations of pre-reionization stars; these are sub-halos
that just exceeded $\vcritr$ at $\zrei$ but have low enough
values of $\vcirc(\zsat)$ that their post-reionization star
formation was strongly suppressed.
If the pre-reionization threshold at $\vcritr$ were smooth
rather than sharp, then some additional fainter systems might have
significant fractions of pre-reionization stars.
However, the general conclusion that pre-reionization star 
formation should be a small fractional contribution at all satellite
luminosities seems fairly robust, provided this star formation
is suppressed in halos below the atomic cooling threshold, as
seems to be required to match the observed luminosity distribution.

Figure~\ref{fig:lf_nosel} shows the complete stellar luminosity function of 
MW satellites inside 400 kpc, in the absence of
any selection effects or incompleteness, again for the fiducial model.
(We choose 400 kpc for ease of comparison to \citealt{tollerud08}.)
In contrast to other figures, it shows the luminosity function
for the whole sky (4$\pi$ sr) and in terms of $dN/dM_V$ (i.e., in bins
of 1 magnitude). Absent selection effects,
the luminosity function continues to rise toward faint magnitudes
\citep[as noted by][]{lfpaper}, contrary to the almost flat luminosity
distribution of observed dwarfs. The total number of satellites within 400~kpc
brighter than $M_V=0$ expected for the fiducial Model 3B is 230$\pm$ 35.
This value is somewhat lower than the 400 derived by \citet{tollerud08},
but since both estimates extrapolate the number of known dwarfs by
a factor of $\sim 10$, we do not place much weight on this difference.

None of the models shown in 
Figures~\ref{fig:hypo1_mv}--\ref{fig:gnedin_combinations}
reproduce the brightest observed bin --- i.e., they all fail
to produce satellites as bright as the SMC and the LMC.  
Our successful models have low stellar mass fractions, $F_* \sim 10^{-3}$,
even well above the photo-ionization threshold $\vcrit$.
The most massive sub-halos in our Monte Carlo realizations have
typical mass $\Msat \sim 10^{11} M_\odot$ (ranging from
$10^{10.5} M_\odot$ to $10^{11.4} M_\odot$), with second-ranked
halos that are $0.2-0.4$ dex less massive.
Reproducing the $\sim 10^9 M_\odot$ stellar masses of the
Magellanic Clouds then requires much higher stellar mass 
fractions $F_* \sim 0.05$.  To reproduce the full satellite
population, the efficiency of gas accretion and star formation 
must continue to rise with halo mass above $\vcrit$, or at least it
must be higher for the SMC and LMC hosts. Since the number of bright
SMC and LMC-like objects in our model are determined mainly by one
parameter $F_*$(because these objects are not suppresed by the 
photo-ionization), that rise of star formation efficiency can not
be accomodated with our simple model without introducing additional
parameters.

\subsection{Velocity dispersions, central masses, and radial
distributions}\label{sec:veldisp_radial}

\begin{figure*}
\includegraphics[width=\textwidth]{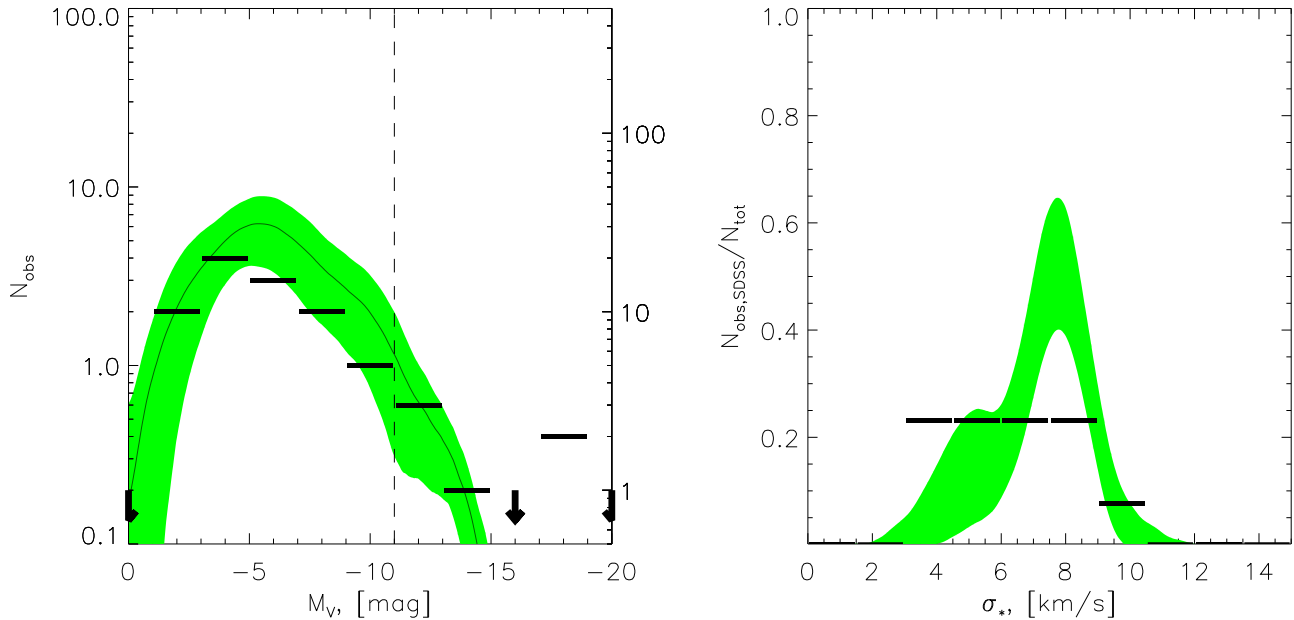}
\caption{Predictions of Model~3B with the fiducial parameters
$(F_*,\vcrit,\zrei,\vcritr)=(10^{-3},35\kms,11,10\kms)$ compared 
to the observed distributions of absolute magnitude (left) and
stellar velocity dispersions (right).
The format of the left panel is the same as Figure~\ref{fig:hypo1_mv}.
The right panel shows predicted and observed velocity dispersions
only for the SDSS dwarfs --- i.e., those with $M_V > -11$ --- with
data taken from \cite{simon07}.
}
\label{good}
\end{figure*}

As discussed in \S\ref{sec:disp_modeling},
predicting stellar velocity dispersions requires 
assumptions beyond those needed to compute $\Nobs$.
In particular, we assume that the satellites' host sub-halos
have NFW profiles with concentration given by the theoretically
expected mean $c(M)$ relation at $\zsat$, and that subsequent
dynamical evolution (e.g., tidal stripping) does not alter the mass
distribution of the inner parts of the sub-halo probed by the stars.
We also take the observed stellar radii ($20-150$ pc, see
\S~\ref{sec:disp_modeling} for details) as input rather than predicting
them from a physical model.  
With these assumptions,
the right panel of Figure~\ref{good} shows
the predicted distribution of stellar velocity dispersions for Model~3B
with our fiducial parameter choices.
The characteristic value and narrow spread of velocity dispersions for the
newly discovered SDSS dwarfs arises quite naturally from these models,
despite the large range of stellar luminosities and host sub-halo masses. 
The predicted distribution is more sharply peaked than the observed one,
probably because we did not include scatter in the halo concentration-mass
relation and did not include observational uncertainties in the dispersion
measurements.  The mean value of $\sigma_*$ differs by $<$ 20\% between
data and model, but we consider this small 
discrepancy is not worrisome, given the simplicity of our dynamical
modeling.

\begin{figure}
\includegraphics[width=0.5\textwidth]{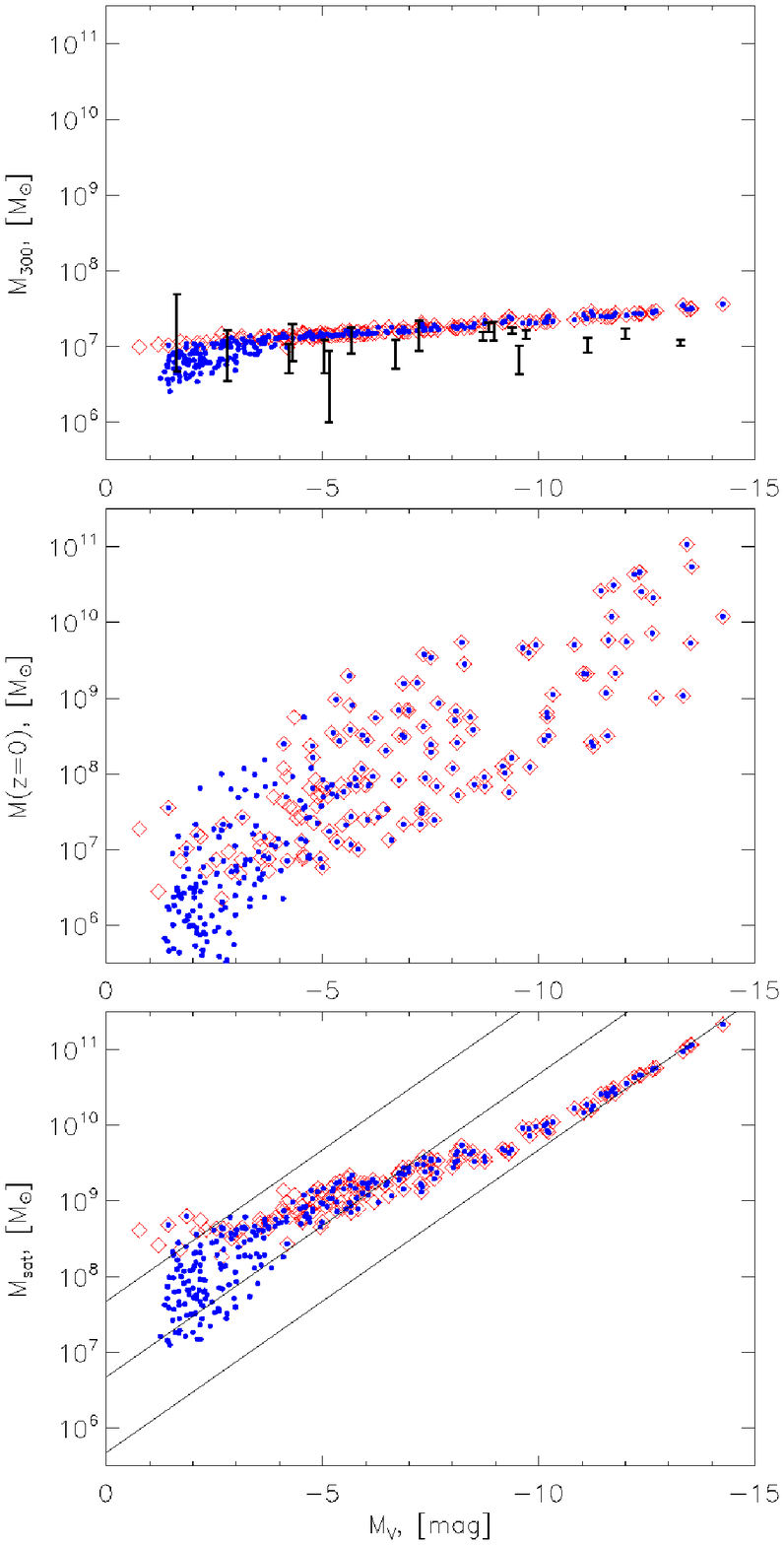}
\caption{
Masses of the DM sub-halos within the central 300 pc (top), their total
present-day masses (middle) and
their masses at the time of accretion into larger halos (bottom). We only 
show halos hosting
observable satellites within the MW virial radius, as a function
of satellite luminosity.
Red diamonds show all the observable galaxies from six
realizations of the fiducial Model 3B with
$(F_*, \vcrit, \zrei, \vcritr)=(10^{-3}, 35\kms, 11, 10\kms)$. 
Blue filled circles show the predictions of Model~3A,
which includes pre-reionization dwarfs (or, equivalently, has $\vcritr = 0$).
Error bars show the estimates of $\mthree$ for observed MW satellites
from~\citep{strigari08}. Solid lines in the bottom panel show,
from top to bottom,
$M_*/\msat=10^{-5}$, $10^{-4}$, and $10^{-3}$.
Our models do not incorporate scatter in the concentration-mass relation;
adding the theoretically expected scatter 
would add roughly 0.15 dex of rms scatter to the $\mthree$ predictions.
}
\label{fig:m300}
\end{figure}

The total masses of dwarf satellites are difficult to determine
observationally because of the small extent of the stellar distributions
relative to the expected extent of the dark matter sub-halo.
However, \cite{strigari08} show that the total mass (principally
dark matter) within a radius of 300 pc, $\mthree$, can be inferred robustly
from observations for nearly all of the known satellites.
The top panel of Figure~\ref{fig:m300} compares the fiducial
model predictions of $\mthree$ to the \cite{strigari08} measurements.
The model (red diamonds)
naturally reproduces the key result of \cite{strigari08}:
over an enormous range of luminosities, the satellites have a 
narrow range of $\mthree$, tightly concentrated around $10^7 M_\odot$.
The theoretical prediction is artificially tight because we have
not included scatter in halo concentrations, which would produce
roughly 0.15 dex (rms) of scatter in $\mthree$ (see \citealt{maccio08},
figure 1).  The model predicts a weak trend of $\mthree$ with 
luminosity, which is not evident in the data (but is similar to
that predicted by \citealt{maccio08}).

While the $\mthree$ range of the satellites is low, the range of
{\it total} sub-halo masses (at $z=0$) is more than three orders
of magnitude, as shown in the middle panel of Figure~\ref{fig:m300}.
The trend of total mass with luminosity is much stronger than the
trend for $\mthree$, though there is a large scatter in mass at
fixed luminosity because of tidal stripping.
The near constancy of $\mthree$ is a consequence of the
density profiles of CDM halos: NFW halos with the theoretically
predicted $c(M)$ relation have only a weak dependence of 
$\mthree$ on total mass over the range $\sim 10^7 - 10^{10} M_\odot$
that hosts observed Milky Way satellites (see \citealt{maccio08}
for further discussion).  Thus our models and the models of 
\cite{maccio08} are able to reproduce the narrow
observed range of $\mthree$ without much difficulty
(see also \citealt{li08}, who examine $M_{\rm 600}$ rather than $\mthree$).
We note, however, that if we also allow satellites to form stars
with efficiency $F_* = 10^{-3}$ before reionization (Model~3A),
then the $\mthree$
range for the lowest luminosity dwarfs, with $M_V > -3$, extends
downwards to $\mthree \sim 10^{6.5} M_\odot$ 
(blue circles in Figure~\ref{fig:m300}).
Thus, careful
dynamical measurements for the faintest dwarfs could in principle
distinguish whether they arise mainly from pre-reionization
star formation or from highly suppressed post-reionization star
formation in more massive halos. 
It is noticeable that our model as well as the models of
\citet{maccio08} and \citet{li08} predicts that 
$\mthree$ or $M_{\rm 600}$ should slightly
increase with galaxy luminosity contradicting the observations,
where there is no correlation at all of $\mthree$ versus
luminosity\citep{strigari08}. The reason of this disagreement is
yet to be understood. It either can be caused by some problems with 
the data (selection effects or systematics in $\mthree$ measurements)
or by some astrophysical effects. For example \citet{maccio08}
eliminates the correlation of $\mthree$ versus luminosity by
assuming that the inner profile of the halos with low concentration
(i.e. massive halos) is modified during the process of tidal
stripping\citep{kazantzidis04}.

Comparing the middle and upper panels shows that a small number
of objects have $M(z=0)$ lower than $\mthree$, which is possible
because we calculate $\mthree$ based on the sub-halo profile
at accretion.  The tidal radii of these systems are $< 300$ pc,
but they are all faint satellites for which the stellar Plummer
radii are small.  While their true $\mthree$ values should be
$M(z=0)$, the values calculated in the upper panel are
probably more directly comparable to the quantities estimated
by \cite{strigari08}, who extrapolate to 300 pc for the faintest
systems assuming that they are not tidally truncated within this radius.
To minimize the tidal effects one may also compute the masses within 
100 pc instead of 300 pc. For our simulated galaxies we also derive 
$M_{\rm 100}$, which are in the range $1\times10^6-4\times10^6 \msun$
and are also
consistent with the $M_{\rm 100}\approx 1\times 10^6-3\times10^6\msun$
measurements from  \citet{strigari08}(supplementary information).

The bottom panel of Fig.~\ref{fig:m300} shows
the value of 
$\msat$ as a function of luminosity. The relation obviously reflects the
underlying formula used to assign stellar masses to the DM halos
(eqn.~\ref{eq:gnedin_recipe_vc}), and the scatter caused by
the range of accretion redshifts (which affects the $\Msat-\vcirc$
mapping) is small.
Even the faintest observable dwarfs have 
$\Msat \sim 10^{8.5} M_\odot$, but they have star formation
efficiencies of only $\sim 10^{-5}$.
The difference between the middle and bottom
panels illustrates the effect of tidal stripping. 
Nearly all the spread of $M(z=0)$ at fixed $M_V$ comes from different 
degree of tidal stripping.

\begin{figure}
 \includegraphics[width=0.5\textwidth]{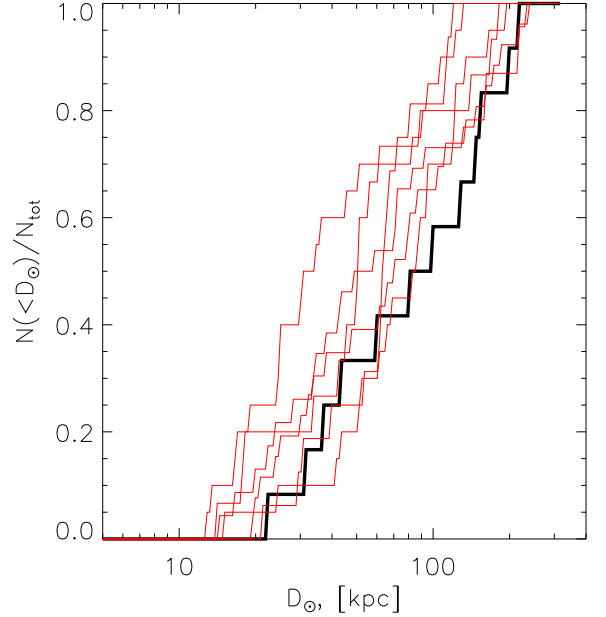}
\caption{Comparison of the model predictions for the 
cumulative distance distribution of the satellite galaxies
with those observed in the SDSS (black line). The
predictions of the Model 3B with $(F_*, \vcrit, \zrei, \vcritr)=(10^{-3},
35\kms, 11, 10\kms)$ are shown as red lines.}
\label{fig:distance_distr}
\end{figure}

Figure~\ref{fig:distance_distr} compares the distribution of
heliocentric distances of the MW satellites found in the SDSS
to the predicted distribution for $M_V > -11$ satellites
from our fiducial model.
We show one distribution for each of the six Monte Carlo 
halo realizations.
There are significant halo-to-halo variations in the predicted
distributions, and the observed distribution follows the lower
envelope of the predictions.
The distance distribution is strongly influenced by the
radial selection effects (the model predictions would be
very different if we did not include them), but it also
depends on the radial profile of sub-halos and the dependence
of this profile on $\Msat$ and $\zsat$, so matching the
observed distribution is a significant additional success
of the model.

\begin{deluxetable}{cccc}
\tablecaption{Satellites used for the
analysis and parameters adopted\label{tab:satellite_list}}
\tablehead{
\colhead{Galaxy} & \colhead{M$_V$} & \colhead{$\sigma_{*}$}
&\colhead{$D_\odot$}\\
\colhead{Name} & \colhead{mag} & \colhead{km/s}
&\colhead{kpc}
}

\startdata
Bootes             &  -6.3 & 6.6  & 60 \\
Canes Venatici II  &  -4.9 & 4.6  & 150\\
Carina             &  -9.4 & 6.8  & 100\\
Coma               &  -4.1 & 4.6  & 45 \\
Canes Venatici I   &  -8.6 & 7.6  & 220\\
Draco              & -8.75 & 10.0 & 80 \\
Fornax             & -13.2 & 10.5 & 138\\
Hercules           &  -6.6 & 5.1  & 130\\
Leo I              & -11.5 & 8.8  & 250\\
Leo II             &  -9.6 & 6.7  & 205\\
Leo IV             &  -5.0 & 3.3  & 160\\
LMC                & -18.6 & -    & 49 \\
Sagittarius        & -12.1 & 11.4 & 24 \\
Sculptor           & -11.1 & 6.6  & 80 \\
Sextans            &  -9.5 & 6.6  & 86 \\
Segue 1            &  -1.5 & 4.3  & 23 \\
SMC                & -17.2 & -    & 58 \\
Ursa Minor         &  -9.0 & 9.3  & 66 \\
Ursa Major I       &  -5.5 & 7.6  & 100\\
Ursa Major II      &  -4.2 & 6.7  & 30 \\
Willman I          &  -2.7 & 4.3  & 40
\enddata
\end{deluxetable}

\section{Conclusions}\label{sec:conclusions}

The satellite discoveries in the SDSS
\citep{willman05,belokurov06,belokurov07,zucker06,irwin07,koposov07,
walsh07} 
have transformed our understanding of the MW's dwarf satellite
population, extending the luminosity range by two orders of magnitude
and the implied number of systems by a factor of 20.
Careful quantification of the SDSS satellite detection efficiency
\citep{lfpaper,walsh08} allows models that specify the relation
between dark matter sub-halos and their stellar content to be
tested quantitatively against the observations.
We have shown that CDM-based models incorporating previously
advocated, physically plausible mechanisms for suppressing
the stellar content of low mass halos can reproduce the observed
properties of the known satellite population, including
their numbers, luminosity distribution, stellar velocity
dispersions, central masses, and heliocentric radius distribution.
However, parameters of these models are tightly constrained, and
alternative assumptions lead to conflict with the data.
In summarizing our results, it is useful to review both
what works and what doesn't.

What works is a model in which the photo-ionizing background
suppresses gas accretion onto halos with 
$\vcirc(\zsat) < \vcrit \approx 35\kms$ \citep{quinn96,thoul96,bullock00},
with the smooth mass-dependent suppression suggested by 
numerical simulations (eqn.~\ref{eq:gnedin_recipe_vc};
\citealt{gnedin00,hoeft06,okamoto08}),
and inefficient molecular cooling (and/or stellar feedback)
drastically reduces the efficiency of star formation in 
pre-reionization halos below the hydrogen atomic line cooling
threshold $\vcritr \approx 10\kms$
\citep{haiman97,barkana99,machacek01,wise07,oshea08,bovill08}. 
There is some degeneracy between this model's two main parameters,
$\vcrit$ and $F_*$, as shown in Figure~\ref{fig:gnedin_combinations},
but with either parameter fixed the other is fairly well
constrained (Figure~\ref{fig:gnedin_mv}).
The other two parameters, $\zrei$ and $\vcritr$, just need to be 
in a range that keeps pre-reionization star formation too low
to affect the observable luminosity function.
For the values $\vcrit = 25-35\kms$ favored by numerical
simulations, $F_*$ must be $\la 10^{-3}$, so even sub-halos
above the $\vcrit$ threshold have star formation efficiency
far lower than the values $F_* \approx 0.1 - 0.4$ found for
bright galaxies (e.g., 
\citealt{pizagno05,mandelbaum06,dutton07,gnedin07,xue08}).

If we assign stellar extents based on observations, and make the
reasonable dynamical assumptions discussed in 
\S\ref{sec:disp_modeling}, then our fiducial model naturally explains
the characteristic value and narrow spread of stellar velocity
dispersions found for SDSS dwarfs by \cite{simon07}.
It also explains the characteristic value and narrow range of
$\mthree$ values found by \cite{strigari08}.
The $\mthree$ values do not depend on the assumed stellar extent,
and their narrow range arises from the theoretically predicted
structure of CDM halos, which have a weak dependence of $\mthree$
on total halo mass over the range 
$M_{\rm halo} \sim 10^8 - 10^{11} M_\odot$.
Thus any CDM-based model that prevents formation of observable
dwarfs in halos below $\sim 10^7 M_\odot$ should
qualitatively reproduce the Strigari et al.\ (2007, 2008)
results (e.g., \citealt{li08,maccio08}).
Tempering this success, however, is the fact that 
the total $z=0$ sub-halo masses in our model
span three orders of magnitude; some of this range is a consequence
of tidal stripping, but the span of $\msat$ values is only
slightly narrower.
The model, in combination with the radial selection biases
found by \cite{lfpaper}, also explains the observed heliocentric
radius distribution of the SDSS dwarfs, which tests the predicted
Galactocentric radius distribution of subhalos and its dependence
on mass and accretion redshift.

Many alternative models fail badly in reproducing the observed
luminosity distribution.  Models with constant $M_*/\msat$ predict
far too many faint satellites relative to bright satellites.
The SDSS discoveries and luminosity-dependent selection biases
do not in themselves resolve the ``missing satellite'' discrepancy
highlighted by \cite{klypin99} and \cite{moore99};
strong mass-dependent suppression of star formation efficiency
is still required to reconcile CDM predictions with observations.
A simple model in which 
$M_*/\msat = 10^{-3} (\Omega_b/\Omega_m) (\Msat/10^{10} M_\odot)^2$
for $\msat < 10^{10} M_\odot$ is reasonably successful at matching
the observations.  This successful ``empirical''
model has a mass dependence of star formation efficiency roughly like
that of the successful, physically motivated photo-ionization model
(eqn.~\ref{eq:gnedin_recipe_vc}; note that $\msat \propto \vcirc^3$ at 
fixed $\zsat$).

Models with {\it sharp} suppression of star formation below the
photo-ionization threshold $\vcrit$ fail at intermediate luminosities,
$M_V \sim -8$.  Pre-reionization star formation can provide the
population of faint dwarfs in such a model, 
but there is an unacceptable gap between the
faint and bright populations (or, for parameter choices that fill
the gap, there is an excess of dwarfs at other luminosities).
It is striking, therefore, that the {\it form} of the 
mass-dependent photo-ionization suppression found in numerical
simulations is just that required to match the shape of the
observed luminosity distribution.
However, the conversion of accreted baryons to stars must
be very inefficient for our fiducial model to work, and it
is not obvious why this conversion efficiency should be
mass independent.

The most interesting of our ``negative'' conclusions is that star
formation in halos before reionization must be extremely inefficient
to avoid producing too many satellites in the range $0 \ga M_V \ga -6$.
Examination of Figure~\ref{fig:gnedin_mv} suggests that the 
upper limit on the fraction of halo baryons converted to stars is
a few~$\times 10^{-4}$ for $\zrei=11$, or $10^{-3}$ if reionization
is pushed back to $\zrei=14$.  \cite{madau08} have reached exactly
the same conclusion, with a similar numerical value for the 
efficiency limit, using the {\it Via Lactea II} simulation instead
of a semi-analytic method to predict the model sub-halo population.
Suppression of star formation in halos below the hydrogen atomic
line cooling threshold is physically plausible, as the metallicity
is low and molecular cooling should be inefficient.
For agreement with $\Nobs$, we require pre-reionization suppression
in halos with $\vcirc(\zrei) < \vcritr \approx 10\kms$.

There are several caveats to these conclusions.  First, as discussed
in \S\ref{sec:observed_distribution},
reproducing the Magellanic Clouds requires that the most massive
sub-halos have $M_*/\msat \sim 0.05-0.1$, well above the 
$F_* \sim 10^{-3}$ of our fiducial model.
Thus, the photo-ionization suppression described by 
equation~(\ref{eq:gnedin_recipe_vc}) must join onto a continuing
increase of star formation efficiency with sub-halo mass
above $\vcrit$, an increase that is presumably driven by other physical
mechanisms.
Indeed, there is nothing about our results that necessarily picks
out photo-ionization as the suppression mechanism in low mass sub-halos,
but it is a mechanism that comes in naturally (one might argue
inevitably) at the desired scale \citep{bullock00},
and the numerically calibrated form yields
a good match to the observed luminosity distribution.

In our fiducial model, even the faintest SDSS dwarfs form most of their
stars after reionization, but they have $\vcirc(\zsat)$ far enough
below $\vcrit$ that their star formation is highly suppressed 
according to equation~(\ref{eq:gnedin_recipe_vc}).
The SDSS dwarfs are physically a continuum with the classical
dwarfs, and their much lower luminosities are a consequence of
the highly non-linear relation between star formation efficiency
and halo mass below $\vcrit$.
Halos with $\vcirc(\zrei) > \vcritr$ form pre-reionization stars,
but in nearly all cases they grow large enough by $\zsat$ that
the post-reionization population dominates by a large factor.
A small number of systems with $M_V \approx -5$ could have large
fractions of pre-reionization stars, but at any luminosity
such systems are rare.
These conclusions are robust within our framework, but
if we allowed for departures from our adopted prescriptions ---
in particular if photo-ionization suppression for $\vcirc \ll \vcrit$ 
were more aggressive than equation~(\ref{eq:gnedin_recipe_vc}) implies and
pre-reionization suppression weaker than we have assumed ---
then it might be possible to construct models in which many dwarfs
with $M_V \ga -6$ are pre-reionization ``fossils.'' 
The efficiency of converting halo baryons to stars
in these systems must still be $\sim 10^{-4}$ or less to avoid
producing too many faint satellites.
\cite{bovill08} and \cite{salvadori09} have argued that halos
cooling by $H_2$ before reionization naturally give rise to the
physical and chemical properties of the SDSS dwarfs.
However, even the low star formation efficiencies $\sim 0.5\%-2\%$ found by
\cite{salvadori09} appear far too high to be consistent with the
observed number counts. On the other hand, \citet{busha09} propose a model in
which post-reionization suppression of star formation is highly
efficient (a sharp threshold) but the star formation efficiency
in pre-reionization halos is strongly mass dependent,
effectively spreading the low luminosity peak evident in
our Figure 6 up towards higher luminosities so that it
fills out the entire faint end of the luminosity function.

A third caveat is that we do not explain the origin of the observed
stellar extents; we just show that once the observed extents are
adopted as inputs, then the observed stellar velocity dispersions
emerge naturally.  One possible explanation is that the baryons
in low mass halos condense {\it until} they reach a scale at
which the velocity dispersion is a few $\kms$, and that this
minimum dispersion provides the conditions necessary for star 
formation.  We also have not attempted to explain the chemical 
abundance distributions or star formation histories of the
satellites (see, e.g., \citealt{orban08,salvadori08,salvadori09}).

A final caveat is that we have assumed that all dwarfs luminous enough
to be found in the SDSS also lie above the surface brightness
threshold for detection, which is about 30 mag arcsec$^{-2}$
\citep{lfpaper}.  Since some of the known satellites approach
this threshold, it is possible that others fall below it.
A large population of lower surface brightness dwarfs would
change the number counts that our model reproduces.
Note also that a large population of pre-reionization dwarfs
would be observationally allowed if they lie below the
surface brightness threshold; however, even in this scenario the 
pre-reionization dwarfs do not account for the presently known
satellites.  Deeper large area imaging surveys, such as PanSTARRS,
the Dark Energy Survey, and LSST, will show whether the MW
satellite population includes a significant number of lower surface
brightness systems.

Our model makes several predictions that can be tested by these 
upcoming surveys or by further follow-up studies of known dwarfs.
Deeper surveys should reveal many more satellites, more than 
200 with $M_V < 0$ and $D_\odot < 400\,$kpc over the full sky,
with the luminosity function shown in Figure~\ref{fig:lf_nosel}.
Deep imaging of Andromeda and other nearby galaxies can show
whether they have similar satellite systems, though these searches
will not reach the extremely low luminosities that can be probed 
in the MW.  
Most satellites in our model 
have stellar extents that are substantially smaller than the
present-day tidal radius of their host halo. Tidal tails and
tidal disruption should be rare, an implication that may be
challenged by photometric evidence on
the profiles and shapes of the ultra-faint galaxies, which have
been interpreted as signs of tidal distortion or disruption
(e.g. Martin \etal 2008).
Measurements of the total sub-halo masses of known dwarfs
would provide a powerful test of the model predictions in 
Figure~\ref{fig:m300}, but the small stellar extents may make
such measurements impossible.  Our models predict that satellites
continue to form stars down to $\zsat$ or below, and many observable
systems should have $\zsat = 1-2$ (see Figure~\ref{fig:accretion_history}).
These predictions may be testable with detailed stellar population
modeling.

Our results greatly strengthen the argument
\citep{bullock00,benson02,somerville02,kravtsov04}
that photo-ionization naturally reconciles the CDM-predicted
sub-halo population with the observed dwarf spheroidal population,
thus solving the ``missing satellite problem'' highlighted
by \cite{klypin99} and \cite{moore99}.
The fiducial model presented here offers a detailed, quantitative
resolution of this problem in light of new, greatly improved
observational constraints, while relying on previously postulated
and physically reasonable mechanisms to suppress star formation
in low mass halos.
The MW satellites provide a fabulous
laboratory for studying galaxy formation at the lowest mass scales,
and much remains to be understood about gas cooling, star formation,
feedback, and chemical enrichment in these systems.
These issues provide challenging targets for numerical simulations
and semi-analytic models, whose predictions can be tested against
detailed studies of the dynamics and stellar populations of the
known dwarf satellites and of the many new satellites that will
be revealed by the next generation of sky surveys.

\acknowledgements{
S. K. was supported by the DFG through SFB 439 and by a EARA-EST Marie
Curie Visiting fellowship. J.~Y. is supported by the Harvard College Observatory
under the Donald~H. Menzel fund.
D.~W. acknowledges support from NSF grant AST-0707985 and the
hospitality of the Institut d'Astrophysique de Paris during part of this work.
S.K. acknowledges hospitality from the Kavli Institute for Theoretical
Physics (KITP) Santa Barbara during the workshop 
``Building the Milky Way''.
We thank James Bullock for his helpful comments on the paper and the
anonymous referee for prompt review and constructive comments.

This paper relies heavily on data from the Sloan Digital Sky Survey.
Funding for the SDSS and SDSS-II was provided by the Alfred P. Sloan
Foundation, the Participating Institutions, the National Science Foundation, 
the U.S. Department of Energy, the National Aeronautics and Space 
Administration, the Japanese Monbukagakusho, the Max Planck Society, 
and the Higher Education Funding Council for England. 
The SDSS was managed by the Astrophysical Research
Consortium for the Participating Institutions, which are listed at
the SDSS Web Site, {\tt http://www.sdss.org/}. 
}

\end{document}